\documentclass[10pt,twocolumn]{IEEEtran}

\DeclareMathAlphabet\mathbfcal{OMS}{cmsy}{b}{n} %GP serve per abilitare il grassetto anche per \mathcal

\usepackage[table]{xcolor}
 \usepackage{comment,graphicx}
 \usepackage{subfigure}
 \usepackage{epstopdf}
\usepackage{psfrag}
\usepackage{acronym}
\usepackage{etoolbox}
\usepackage{epsfig}
\usepackage{colortbl,color}
\usepackage{amsmath}
\usepackage{amssymb}
\usepackage{mathabx}
\usepackage{enumerate}
\usepackage{bbm}
\usepackage{algorithm}
\usepackage{algorithmic}
\usepackage{lipsum,multicol}
\usepackage{stfloats}
\usepackage{arydshln} 
\usepackage{amsfonts}
\usepackage{dsfont}
\usepackage{upgreek}
\usepackage{pgfplots} 
\usepackage{soul}
\usepackage{tabularx}
\usepackage{multirow}
\usepackage{graphicx}
\usepackage{pgfplots}

\usepackage{cite}
\usepackage{verbatim}
\usepackage{blkarray}
\usepackage{bm}
\usepackage{tikz}
\usepackage{booktabs}
\newcommand{\quot}[1] {``{{#1}}''}

\definecolor{LightCyan}{rgb}{0.88,1,1}
\definecolor{LightGreen}{rgb}{0.88,1,1}

\acrodef{CLEAN}{CLEAN}
\acrodef{CFAR}{constant false alarm rate}
\acrodef{CFR}{channel frequency response}
\acrodef{CIR}{channel impulse response}
\acrodef{GEM}{ghost-effect mitigation}
\acrodef{HPBW}{Half Power Beam Width}
\acrodef{IDFT}{inverse discrete Fourier transform}
\acrodef{IFBW}{Intermediate Frequency Band Width}
\acrodef{Lidar}{laser-based radar}
\acrodef{MAP}{maximum a posteriori}
\acrodef{MIMO}{multiple-input multiple-output}
\acrodef{NM}{noise-masking}
\acrodef{pdf}{probability density function}
\acrodef{pmf}{probability mass function}
\acrodef{RIS}{reconfigurable intelligent surface}
\acrodef{RMSE}{root-mean-square error}
\acrodef{R-SLAM}{radio-SLAM}
\acrodef{SLAM}{simultaneous localization and mapping}
\acrodef{THz}{Terahertz}
\acrodef{UAV}{unmanned aerial vehicle}
\acrodef{VNA}{Vector Network Analyzer}
\acrodef{V-SLAM}{Visual-SLAM}
\acrodef{5G}{fifth generation}
\acrodef{6G}{sixth generation}
\acrodef{RF}{radio frequency}
\acrodef{psd}{power spectral density}
\acrodef{SFCW}{Stepped-Frequency Continuous-Wave}

\newcommand{\Ncell}{N_{\mathsf{cell}}}
\newcommand{\map}{\mathbf{m}}

\newcommand{\mapk}{\mathbf{m}_k}
\newcommand{\mapi}{{m}_i}
\newcommand{\mapik}{{m}_{i,k}}
\newcommand{\mapsk}{{m}_{1,k}}
\newcommand{\mapek}{{m}_{\Ncell,k}}

\newcommand{\bk}{{b}_{k}}
\newcommand{\bs}{{b}_{0}}
\newcommand{\logoddk}{{\ell}_{k}}
\newcommand{\logoddkprev}{\ell_{k-1}}

\newcommand{\xigem}{\xi_m^{(\mathsf{GEM})}}
\newcommand{\xinm}{\xi^{(\mathsf{NM})}}
\begin{document}
% \title{Towards 6G Integrated Sensing and Localization: Radio SLAM for THz Networks}

\title{Radio SLAM for 6G Systems at THz Frequencies: Design and Experimental Validation}

\author{M. Lotti,~\IEEEmembership{Student Member,~IEEE}, G. Pasolini,~\IEEEmembership{Member,~IEEE}, A. Guerra,~\IEEEmembership{Member,~IEEE}, \\ F. Guidi,~\IEEEmembership{Member,~IEEE}, R. D'Errico,~\IEEEmembership{Senior Member,~IEEE}, and D. Dardari,~\IEEEmembership{Senior Member,~IEEE} %~\IEEEmembership{Member,~IEEE,}% <-this % stops a space
\thanks{
M. Lotti, G. Pasolini (corresponding author), and D. Dardari are with the WiLAB - Department of Electrical and Information Engineering ``Guglielmo Marconi" - CNIT, University of Bologna, Italy. E-mail: (marina.lotti2, gianni.pasolini, davide.dardari)@unibo.it. \\ A. Guerra and F. Guidi are with the National Research Council of Italy, IEIIT-CNR, Italy, and also with the WiLAB. E-mail: (anna.guerra,francesco.guidi)@cnr.it. \\R. D'Errico is with CEA, LETI, MINATEC
Campus, 38054 Grenoble, France, and also with the University of
Grenoble-Alpes, 38000 Grenoble, France. E-mail: raffaele.derrico@cea.fr.   }}

% The paper headers
\markboth{Submitted to IEEE Journal of Selected Topics in Signal Processing}%
{XX \MakeLowercase{\textit{et al.}}: Bare Demo of IEEEtran.cls for IEEE Journals}

\maketitle

%\blue{Alternative titles: \\
%1. Integrated Sensing and Localization at THz: Theoretical Analysis and Experiments.\\
%2. Integrated Sensing and Localization at THz: Theoretical Analysis and Channel Characterization \\
%3. Radio SLAM for THz Networks: Theoretical Analysis and Experimental Validation
%4. Integrated Localization and Sensing at THz: Theoretical Analysis and Experimental Validation}

\begin{abstract}

Next-generation wireless networks will see the convergence of communication and sensing, also exploiting the availability of large bandwidths in the \ac{THz} spectrum and electrically large antenna arrays on handheld devices. In particular, it is envisaged that user devices will be able to automatically scan their surroundings by steering a very narrow antenna beam and collecting echoes reflected by objects and walls to derive a map of indoors and infer users' trajectories using \ac{SLAM} techniques.
In this paper, we address this scenario by proposing original \ac{R-SLAM}  algorithms, derived from image processing techniques, to map the environment and pinpoint the device position in the map starting from measurements sensed by a mobile \ac{THz} radar. %, thus moving from \ac{SLAM} to \ac{R-SLAM}.
Initially, to fully understand the \ac{THz} backscattering phenomenon, we provide an experimental characterization of the \ac{THz} backscattering channel in indoor environments.
Then, the performance of the proposed algorithms is assessed using real-world \ac{THz} radar measurements and is compared with state-of-the-art \ac{SLAM} techniques, demonstrating the superiority of the proposed approaches. 

% The next \ac{6G}  wireless systems demand high data rates, high-resolution content transmissions, and seamless connectivity to enable new applications and scenarios. Thanks to the large available bandwidth, \ac{THz} technology represents a promising game-changer to meet the consequent capacity requirements as well as the integration of sensing capabilities into the same device used for communication. From a localization perspective, the large \ac{THz} bandwidth, typically exploited for communications, allows to attain high range and angular resolution through the realization of large antenna arrays, essential for enabling high-resolution map reconstruction and high-accuracy pose estimation. Thus, \ac{THz} \ac{SLAM} seems a viable alternative to current vision-based systems, which suffer in scarce visibility conditions, and to low-frequency radars, which usually do not allow to achieve precise angular and range accuracy. Undoubtedly, \ac{THz} \ac{SLAM} represents a promising solution to fill such gap, but several aspects have not been investigated yet, like \ac{THz} propagation backscattering characterisation,. 
% This paper aims to provide a unified theoretical and experimental framework for \ac{R-SLAM} by proposing a low-complex Fourier-Mellin–inspired algorithm for pose estimation and automatic mapping and by a delay–angular characterization of the THz channel through extensive indoor measurement campaigns.
\end{abstract}
\acresetall
%\blue{JSTSP - Instructions: \\Prepare a PDF file containing your manuscript in double-column, single-spaced format using a font size of 10 points or larger, having a margin of at least 1 inch on all sides. For a regular paper, the manuscript may not exceed 13 double-column pages, including title; names of authors and their complete contact information; abstract; text; all images, figures and tables, appendices and proofs; and all references. Upload this version of the manuscript as a PDF file "double.pdf" to the ScholarOneManuscripts site. You are encouraged to also submit a single-column, double-spaced version (11 point font or larger), but page length restrictions will be determined by the double-column version.}

% Note that keywords are not normally used for peerreview papers.

\bstctlcite{IEEEexample:BSTcontrol}

\IEEEpeerreviewmaketitle

%\tableofcontents

\section{Introduction}

\IEEEPARstart{N}{}ext-generation cellular networks are credited to becoming the point of convergence of communication and sensing thanks to the availability of high-frequency technologies that will foster the creation of an ecosystem of applications and services exploiting seamless connectivity and data rates at an unprecedented scale \cite{SarEtAl:J20,RajEtAl:J20}. 
%
%{N}{}ext-generation wireless communications networks are celebrated to become an appointment between communication, localization, and sensing, thanks to the support of high-frequency technologies that will propel the creation of an ecosystem of applications and services that will take advantage of seamless connectivity and data rates at an unprecedented scale \cite{SarEtAl:J20,RajEtAl:J20}. 
%
In fact, unlike today's \ac{5G} networks, which are primarily designed for wireless communications, it is expected that \ac{6G} networks will entail a quantum leap towards the integration of sensing capabilities in handheld devices so that the latter will be fully aware of their surrounding 3D environment. 
%
%access a 3D mapping of the radio environment \cite{SaaBenChe:J20,WilBraVis:J21,WanEtAl:C22,LiyEtAl:J21,Del:21}. 
%
In this regard, the ability to operate in the \ac{THz} band is expected to play a crucial role, as the wide bandwidth available will result in high spatial resolution and the feasibility of large antenna arrays will enable unprecedented angular resolution
\cite{AKYILDIZ201416,AKYILDIZ201646, AkyJorHan:J14,YanEtAl:22,J:IEEE_Networks22}.

Traditionally, the \ac{THz} band, namely the frequencies in the range $0.1-10\,$THz, has represented the  \emph{last gap} between radio and optical signals \cite{ChaEtAl:J22,HanEtAl:J22}. 
%
% If in the past, the \ac{THz} gap was due to the inability to realize efficient hardware (e.g., transceivers, antennas) at such frequencies, now the situation has changed as we are experiencing the availability of a more mature technology. Thanks to this technological improvement, new applications and services become possible that require close integration of sensing capabilities with communications.
%
In the past, the \ac{THz} gap was due to the difficulty of realizing hardware (e.g., transceivers, antennas) capable of operating with an adequate performance at these frequencies. However, the situation is changing rapidly as technological problems are being overcome. New applications and services that require tight integration of sensing and communications capabilities are thus becoming possible \cite{ElbMisCha:J21,CheEtAl:J22}.
%
%\red{Infrastructure-less localization/mapping}
% 
In this perspective, mobile devices will become exceptional media for environmental monitoring thanks to their pervasive diffusion and the increasing number of embedded sensors \cite{PasEtAl:J20}. But there is more than that: in fact, according to the \ac{6G} vision, handheld devices are expected to revolutionize our perception of indoor environments and our ability to move inside them, creating digital maps of our surroundings autonomously and pinpointing our position within the map, without the need for a dedicated positioning infrastructure \cite{PasEtAl:J20}. This perspective motivated our work, which proposes original \ac{SLAM} algorithms working on THz signals and evaluates their performance when fed with real measurements taken in an indoor scenario.
% The process of creating digital maps of the world began a couple of decades ago in outdoor scenarios, with the goal of obtaining digitized road maps for vehicle navigation systems.
% It is a fact, however, that nowadays the coverage of indoor maps (that is, of GPS-denied environments) is far less than that of outdoor environments. The reason is easy to understand: the need for the complex and bulky devices currently used for indoor mapping prevents their large-scale adoption.  In fact, nowadays, indoor mapping is mainly carried out using \acp{Lidar} or
% \ac{V-SLAM}  cameras \cite{CadEtAl:J16,DurBai:J06,thrun2006graph,ZelQuiSti:J17,FuEtAl:J22}. Unfortunately, such solutions are very sensitive to lighting conditions, as they require perfect visibility and, in addition, professional \acp{Lidar} are usually expensive, power-hungry and their use is restricted to experienced personnel for security reasons. 
% The starting point is the current situation, where simultaneous localization and mapping requires complex and bulky devices, operated by trained personnel, that incorporate \acp{Lidar} or \ac{V-SLAM}  cameras \cite{CadEtAl:J16,DurBai:J06,thrun2006graph,ZelQuiSti:J17,FuEtAl:J22}. Unfortunately, these devices are not only expensive and power-hungry, but also require perfect visibility, so they are not suitable for incorporation into mobile devices.
%

Currently, accurate \ac{SLAM} is practically implemented by means of complex and bulky devices, operated by trained personnel, that incorporate \acp{Lidar} or \ac{V-SLAM}  cameras \cite{CadEtAl:J16,DurBai:J06,thrun2006graph,ZelQuiSti:J17,FuEtAl:J22}. Unfortunately, these devices are not only expensive and power-hungry, but also require perfect visibility and must be operated manually, so they are not suitable for incorporation into mobile devices tasked with automatically exploring the environment.

%\emph{To overcome these limitations, techniques have been proposed that use radio signals to at least localize the mobile users. However, these solutions pose two problems: the need to install dedicated infrastructures with reference nodes placed in appropriate locations, and lower localization accuracy than that provided by \acp{Lidar}.    
%
%To take this a step further, some studies have proposed improving localization accuracy by treating radio echoes as if they were generated by \quot{virtual} reference nodes in addition to actual reference nodes \cite{WitEtAl:J16}. To avoid the need for a dedicated infrastructure, other studies addressed the exploitation of existing radio frequency signals as sources of opportunity (Wi-Fi Slam, Deep Map), but with scarce resolution and accuracy.} \red{(GP: servono citazioni su questo tema. Tuttavia suggerisco di eliminare l'intero paragrafo che ho messo in corsivo perché spezza il filo logico incentrato sullo SLAM.)}

% A methodological and technological shift was instead proposed in \cite{GuiGueDar:J15,GuiEtAl:J18} through the introduction of the {\it personal radar} concept, 

In this regard, a methodological and technological shift was proposed in \cite{GuiGueDar:J15,GuiEtAl:J18} to address the SLAM problem through the introduction of the {\it personal radar} concept, which concerns the adoption of  \ac{R-SLAM} techniques in handheld devices.
Indeed, taking advantage of the fact that evolutionary \ac{5G} and beyond-\ac{5G} scenarios envisage the integration of quasi-pencil beam antennas at mm-wave and \ac{THz} bands, one can think to use the same hardware to implement (personal) radar capabilities,
overcoming the  limitations of Lidar-based and Visual-based \ac{SLAM} and taking a significant step in the direction of the \quot{integrated communication and sensing} \cite{Li2022} paradigm.

According to the {\it personal radar} concept, it is thus envisioned that mobile devices will be able to accurately scan the environment by transmitting probe radio signals via the generation of narrow radio beams pointing in different directions (beamsteering), and by receiving the signal reflected by the surroundings \cite{GuiEtAl:J18}. Processing such radio echoes will allow the mobile device to retrieve ranging and bearing information, and will make it possible to derive the maps of indoor spaces \cite{PasEtAl:J20,LotEtAl:C21,barneto2022millimeter}.
%This concept is enabled by the joint use of millimeter-wave and massive arrays technologies, thus providing users with a local map of the environment through the precise collection of radio echoes reflected by walls and objects enhanced by near pencil-beam radio scanning capabilities.
%
Jointly with the generation of the map, the self-localization of the device in the actual environment is also carried out by the \emph{personal radar}, which thus turns into an infrastructure-free, zero-cost, non-intrusive, and accurate indoor localization system based on \ac{R-SLAM} techniques. 

So far, state-of-the-art \ac{SLAM} algorithms have been designed primarily for \acp{Lidar}, which, for each complete scan (i.e., for all angular directions), output a scan vector where the information on the distance (range) from the laser source of only one object (if any) is included for each angular direction (resulting in a \emph{one-dimensional} representation of the environment). Contrarily, radar measurements are characterized by a richer information content than those of \acp{Lidar}. 
In fact, for each complete scan, radars generate a \emph{two-dimensional} representation of the environment, namely, the range-angle matrix, which can ultimately be meant as an image of the surrounding scenario.
Indeed, considering the high range-angle resolution expected from THz radars, radio images are expected to be a fairly accurate representation of the surrounding environment. 

This opens the door to the adoption for \ac{R-SLAM} purposes of algorithms typically designed for image processing, like those based on Fourier-Mellin transforms \cite{QinDefDec:J94,RedCha:J96}, which (with appropriate adjustments) are expected to better exploit the entire information content of radar images than algorithms working on one-dimensional scan vectors such as the widely-used laser Scan Matching algorithm \cite{HeKoRaAn:C16}. 

Along this direction, in this paper 
%In this paper, motivated by the possibility of integrating the peculiarities of image processing algorithms into \ac{THz} signal processing schemes, we 
we first introduce a Fourier-Mellin-based approach for joint localization and mapping at \ac{THz}, also proposing an ad-hoc simplified version tailored for portable devices, where low-complexity is an important requirement.  Then, the performance of the proposed schemes is assessed using real-world \ac{THz} radar measurements and is compared with state-of-the-art SLAM techniques, demonstrating the superiority of the proposed approaches. To fully understand the \ac{THz} backscattering phenomenon, we also provide an experimental characterization of the \ac{THz} backscattering channel using the same radar.

\subsection{Related Works and Proposed Contribution}
  \label{Sec.Related_Works}
%The papers most directly related to this work have been mentioned in the Introduction and will also be cited in the following sections. 
%Below, we provide an overview of the documents that represent the general background of our work. 

Surveys of the general \ac{SLAM} problem can be found in \cite{CadEtAl:J16,CheEtAl:J22},  %which discuss the way data are associated, maps are represented, predictions are computed, states are represented, 
including (but not limited to) techniques like FastSLAM, GraphSLAM and belief propagation \ac{SLAM} \cite{thrun2006graph,LeiEtAl:J19,MulEtAl:J11,MonEtAl:J02}.

% Due to all these practical and economic reasons, making extensive mapping of indoors with such technologies appears still infeasible.  
%\red{Example of Personal Radar Application}\\
%Indeed, the pervasive diffusion of billions of mobile personal devices will impressively extend the availability of indoor maps with respect to the level that would be achieved using dedicated devices (e.g., robots with laser scanner). 
%Unfortunately, in those preliminary studies, 

%\red{Current Radio-SLAM algorithms}\\
\Ac{R-SLAM}, initially relegated to niche radar applications \cite{CheEtAl:J10}, has recently attracted a lot of interest for its utilization in 5G and next-generation wireless systems in  automotive and mobile personal indoor applications.
In  \cite{Hol19}, an algorithm using the iterative closest point (ICP) graph method to match consecutive scans obtained from an automotive frequency modulated continuous wave (FMCW) radar and odometry is proposed. It exhibits a mean translational error as low as $0.62\,\mathrm{m}$ using  a real-world dataset at millimeter waves collected in large outdoor areas. 
A similar graph-based approach is used in \cite{Hon21} to perform outdoor \ac{SLAM} in various weather conditions. 

Regarding indoor applications based on 5G-like millimeter wave signals, the typical approach is to localize and map by detecting and tracking the dominant multipath components using Bayesian filtering approaches \cite{WitMeiLeiSheGusTufHanDarMolConWin:J16,Ala19,KimEtAl:J20,WymSec:C20,GeEtAl:J20,BarEtAl:J21}. 
In \cite{BarEtAl:J21-2} a full duplex radio/radar technology is proposed for joint communication and sensing. An approach  based on  fingerprinting can be found in \cite{BeaHyu:J19}%,GeEtAl:J20} %KimEtAl:C20}.
, whereas the authors in \cite{WymEtAl:J19,YanEtAl:J22} discuss the possibility to realize \ac{R-SLAM}-like applications with next-generation large intelligent surfaces.
%while in \cite{WymSec:C20}, authors include detection probabilities of hypothesized landmarks into the {mm-waves} \ac{SLAM} in order to enhance user self-localization and environment mapping. 

Attempts to merge typical image processing approaches with \ac{R-SLAM} in the microwave bands can be found in \cite{CheEtAl:J10}, where the Fourier-Mellin transform is used to register consecutive radar images obtained through an FMCW technology within an EKF-SLAM framework, and in \cite{CalEtAl:J11}, where the scale-invariant feature transform (SIFT) is adopted to extract trackable features from the radar image which are subsequently matched with features from laser scans.

To further improve angular and range precision, \ac{THz} technology has been mainly investigated for imaging  \cite{StaEtAl:J20,ValEtAl:G21}, and rarely for localization \cite{ZheEtAl:J22}, but only separately.
Recently, a \ac{R-SLAM} system for indoor flying agents at \ac{THz} has been investigated in \cite{BatEtAl:20} which relies on the deployment of passive tags and a synthetic aperture radar capable of millimeter-level localization accuracy.
%
%Recent efforts have also pushed towards their joint use at \ac{THz} \cite{BatEtAl:20} but  without accounting for the high-resolution beamsteering capabilities offered by near-pencil beam \ac{THz} antennas. Indeed, considering joint sensing and localization at \ac{THz} is challenging due to
%the narrow beamwidths \cite{AkyJor:J16}, that guarantee precise angular accuracy but that are also sensible to beam misalignment and are difficult to track.
%
To the authors' knowledge, \ac{R-SLAM} exploiting the imaging potential at \ac{THz}  has yet to be explored and validated experimentally. %\cite{FaiEtAl:J20,CheEtAl:J22}.

Considering the limitations of the current literature on \ac{THz} \ac{R-SLAM}, the main contributions of our work can be summarized as follows.

\begin{itemize}
\item We develop an ad-hoc pre-processing scheme, including \ac{GEM} and \ac{NM} strategies,  to reduce the presence of outliers and artifacts that might appear in radar measurements, thus selecting the most informative range-angle information to be provided to the \ac{R-SLAM} algorithm.
\item We propose an \ac{R-SLAM} algorithm based on the  Fourier-Mellin transform, which is typically used in image processing,  that accounts for the \ac{THz} radar-specific  observation model. A  simplified version of the previous algorithm  is also introduced,  involving less computational complexity and thus  facilitating its integration into personal devices. 
%The proposed solution shows similar accuracy for pose estimation with lower computational complexity. Furthermore, such an \ac{R-SLAM} algorithm also includes an occupancy mapping scheme for map reconstruction that allows a probabilistic reconstruction of the surrounding.
\item We provide a joint delay and angular characterization of the backscattering \ac{THz} channel through an extensive measurement campaign carried out in an  indoor environment. 
%The measurement set-up emulated the radar scanning capabilities thanks to the use of directive antennas and a linear–angular positioner, allowing precise mechanical steering.
To the best of authors' knowledge, even if sub-\ac{THz} and \ac{THz} $1$-way channels have already been modelled for communication purposes \cite{PomDer:J19,LotCaiDer:C22,SerEtAl:J22} and \ac{THz} backscattering measurements have been recently performed in bistatic configurations \cite{AdiEtAl:J22}, the literature still lacks a  characterization and understanding of the \ac{THz}  backscattering channel in a quasi-monostatic configuration for \ac{R-SLAM}-based  applications. 
\item Finally, the performance of the proposed \ac{R-SLAM}  schemes is compared with the state-of-the-art \ac{SLAM} approach for \ac{Lidar} measurements, namely the {\em Laser Scan Matching} algorithm \cite{HeKoRaAn:C16}. It is worth noting that, to the authors' knowledge, no article provides such a comparison with experimental data at \ac{THz}.
%we provide an extensive performance evaluation by integrating real-world measurement data with the pre-processing and \ac{R-SLAM} algorithmic introduced in this paper. 
Using real-world radar measurements, the numerical results show the possibility of realizing {\em mapless} and {\em infrastructure-less} radio-based indoor localization and mapping at \ac{THz} with cm-level and as low as mm-level accuracy. 
\end{itemize}
The rest of the paper is organized as follows: %Sec.~\ref{Sec.Related_Works} provides a survey of the literature on \ac{SLAM} and \ac{R-SLAM} techniques;
Sec.~\ref{Sec:RSLAM} introduces the {R-SLAM} problem, and radar-signal pre-processing schemes including mapping;
Sec.~\ref{Sec.Rel_Pose_Estimates} describes the relative pose estimation algorithms considered and proposed in this paper;  Sec.~\ref{Sec:Measures} illustrates  the measurement campaigns and the delay--angular channel characterization;
Sec.~\ref{sec:slamperformance} discusses the performance of \ac{R-SLAM} algorithms; and Sec.~\ref{sec:conclusions} draws the conclusions \footnote{Part of the content of this manuscript appeared in our conference paper \cite{LotEtAl:C21}, which, however, did not include the original R-SLAM algorithms introduced in this manuscript nor the different scenarios considered here for their experimental validation. This paper also proposes additional results from the backscattering channel characterisation campaign.}.

\subsection{Notation}
% Boldface lower-case letters are vectors (e.g., $\textbf{x}$), whereas boldface capital letters are matrices (e.g., $\textbf{A}$), where $\textbf{I}_{M \times N}$ and $\textbf{0}_{M \times N}$ are respectively the identity and zero matrices of size $M \times N$. 
Boldface lower-case letters are vectors (e.g., $\textbf{x}$), whereas boldface capital letters are matrices (e.g., $\textbf{A}$), $\textbf{I}_{N}$ and $\textbf{0}_{N}$ are respectively the identity and zero matrices of size $N \times N$, while $\textbf{0}_{M \times N}$ is the zero matrix of size $M \times N$. 
The notation $a_{n,m}=[\textbf{A}]_{n,m}$ represents the $(n,m)$th element of matrix $\textbf{A}$.  %$\overset{\curvearrowright}{\mathbf{A}}$ indicates the rotation of matrix $\mathbf{A}$. 
$\left( \cdot \right)^T$ and $\left( \cdot \right)^*$ indicate, respectively,  the transpose and complex conjugate operators.

\begin{figure*}
\centering
    \input{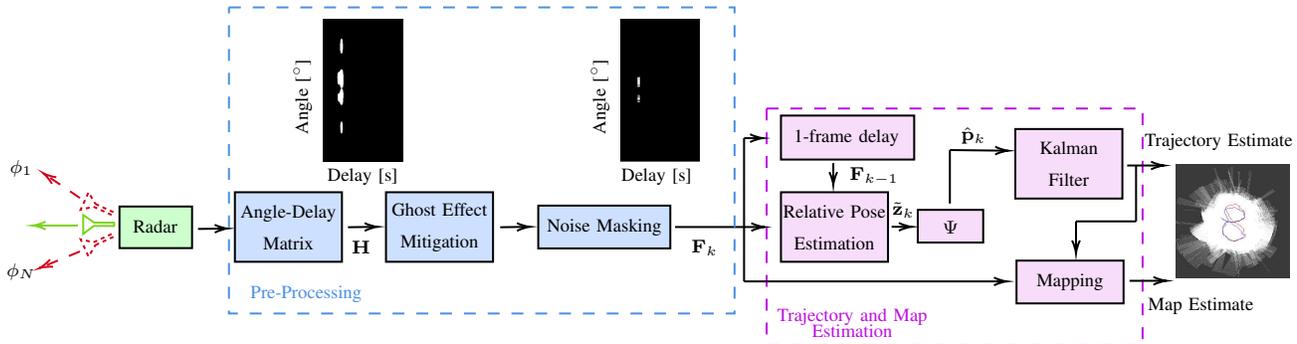}
    \caption{\ac{R-SLAM}: Proposed processing chain. We denoted with $\Psi$ the operations of \eqref{eq:op1}-\eqref{eq:op2} that transform the relative pose ${\tilde{\mathbf{z}}_k=[\tilde{dx}, \tilde{dy}, \tilde{d\theta}]^T}$ on the absolute pose $\hat{\mathbf{p}}_k$. Possible implementations of the ``Pose Estimation" block are depicted in Figs.~\ref{fig:Fourier_Mellin_Processing}-\ref{fig:Approx_Fourier_Mellin_Processing}.}
    \label{fig:Processing_chain}
\end{figure*}

\section{THz Radio SLAM} \label{Sec:RSLAM}

In this section, we introduce the proposed \ac{R-SLAM} approach tailored to a mobile user equipped with a terminal offering THz-based radar functionalities.  We first formalize the problem statement and then describe the processing chain that, starting with raw radar measurements, leads to the estimation of the user's trajectory and the automatic mapping of the environment.

\subsection{Problem Statement}
We consider a $2$D scenario in which the state of a mobile user at the (discrete) time instant $k$, with time step $T_{\text{F}}$, is denoted by 
\begin{equation} \label{eq:userstate}
\textbf{x}_k=[x_k, y_k, \dot x_k,\dot y_k, \theta_k, \dot \theta_k]^T
\end{equation}
 which accounts for the user's position coordinates $(x_k, y_k)$, orientation $\theta_k$ and their variation speeds $(\dot x_k,\dot y_k)$ and $\dot \theta_k$, respectively. Moreover, we denote by 
\begin{equation} \label{eq:userspose}
{\mathbf{p}}_k=[x_k, y_k, \theta_k, ]^T 
\end{equation}
the absolute pose of the user.

Our objective is to devise a processing chain that, starting from raw measurements provided by the THz radar,  is capable  of estimating in real-time the trajectory of the mobile user up to time $k$, i.e., the sequence of states $\textbf{x}_{[1:k]}$ (and therefore of poses $\mathbf{p}_{[1:k]}$), as well as the map of the surrounding environment. 
Classical solutions to the joint localization and mapping problem belong to the class of \ac{SLAM} algorithms that are widely investigated in the literature in the case where the measurements source is a laser. In our case, however, the final objective must be achieved starting from radio measurements, which requires ad hoc strategies that fall within the far less investigated  \ac{R-SLAM} field.  
%In the next sections, we will introduce and discuss the \ac{R-SLAM} processing chain shown in Fig.~\ref{fig:Processing_chain}, which we designed, implemented and tested with real radar measurements in the 300 GHz band.

\subsection{Pre-Processing of Radar Signals}
\label{Sec:Preprocessing}
Radio signals backscattered by the environment have to
be properly processed in order to infer the map of the
scenario and, simultaneously, track the position of the user.  
In the following, we will discuss each step of the processing chain, which  is depicted in Fig.~\ref{fig:Processing_chain}.   
In order to keep our discussion general, we will not stick on a particular radar technology, but rather we suppose that the radar equipment provides the sampled \acp{CIR} of the two-way channel for a set $\{\phi_1, \phi_2, \ldots ,\phi_N\}$ of $N$ angular directions. 
These can be obtained, for example, using a MIMO radar where the signal emitted by the transmitting antennas and backscattered by the environment is collected by an array of receiving antennas \cite{bookradar2020}. Such outcomes are then processed as described below. How these \acp{CIR} have been obtained in the real measurement setup in the $300$ $\mathrm{GHz}$ band exploited in this work will be explained in Sec.~\ref{Sec:Measures}.

\subsubsection{Generation of the Angle-Delay Matrix}
\label{Section:Angle-Delay_matrix}

As mentioned above, let us assume that the radar sounder outputs $M$  samples, with sampling time $T_{\text{s}}$ (time resolution), of the backscatter \ac{CIR} for $N$ different angles of view $\phi_n$ uniformly distributed in the range $[-90^{\circ}$, $90^{\circ}]$. The magnitude of these samples are gathered into the \textit{Angle-Delay} matrix  
\begin{align}
    \mathbf{H}=\{ |h_{n,m}| \},\,\, n=1,2, \ldots ,N, \,\, m=1,2, \ldots ,M, 
\end{align}
so that the $n$th row of $\mathbf{H}$ contains the magnitudes of $M$ (noisy) samples of the \ac{CIR} in the angular direction $\phi_n$. In particular, for a given angular direction $\phi_n$, $|h_{n,m}|$ refers to the time instant $t_m=T_{\text{min}}+(m-1)\, T_{\text{s}}$, $m=1,2, \ldots, M$,  with $T_{\text{min}}$ denoting the minimum two-way propagation delay which is considered to remove the contribution of the direct coupling between the transmitter and the receiver antennas of the radar. 
%Clearly, the rows of $\mathbf{AI}$ are the estimates of the sampled impulse responses of the channel for each angular direction. 
%
Clearly, by converting propagation delays into distances between the radar and the reflecting objects, the generic element $|h_{n,m}|$ corresponds to a range $d_m=d_{\text{min}}+(m-1)\,d$ in the angular direction $\phi_n$, where $d_{\text{min}}=c\, T_{\text{min}}/2$ is the minimum detection distance, $d=c\, T_{\text{s}}/2$, and $c$ is the speed of light.

% In the absence of impairments\footnote{The presence of impairments, such as noise or spurious channel responses, and possible countermeasures will be discussed in Sec.\ref{Section:clean} and Sec.\ref{Section:cfar}.},  possible peaks in the magnitude of the impulse response occur at twice the propagation delay between the radar and the targets. 

\subsubsection{Ghost Effect Mitigation}
\label{Section:clean}
Ideally, an antenna oriented in a given direction should only receive signals coming from the same direction, i.e., reflected from objects intercepted by the antenna axis. 
This  property is a peculiar feature of \acp{Lidar} because of the extremely narrow laser beam. 
Unfortunately, even though very narrow beams can be realized  at THz frequencies, 
%this condition is almost impossible to achieve in the case of radio transmissions, because of 
 unwanted  side lobes of the antenna's radiation pattern might catch also echoes coming from other directions, then making \ac{R-SLAM} much more challenging than classical \ac{Lidar}-based \ac{SLAM}. 
In fact, the radar might erroneously infer the presence of an object in the direction where the main beam of the antenna is oriented due to echoes coming from other directions, thus showing ghost artifacts \cite{guidi2016joint}. 

\begin{figure*}[t!]
	\centering \subfigure[$\mathbf{H}$ matrix before the \ac{GEM} procedure.]{\includegraphics[trim= {10 0 35 10}, clip, width=0.4\linewidth]{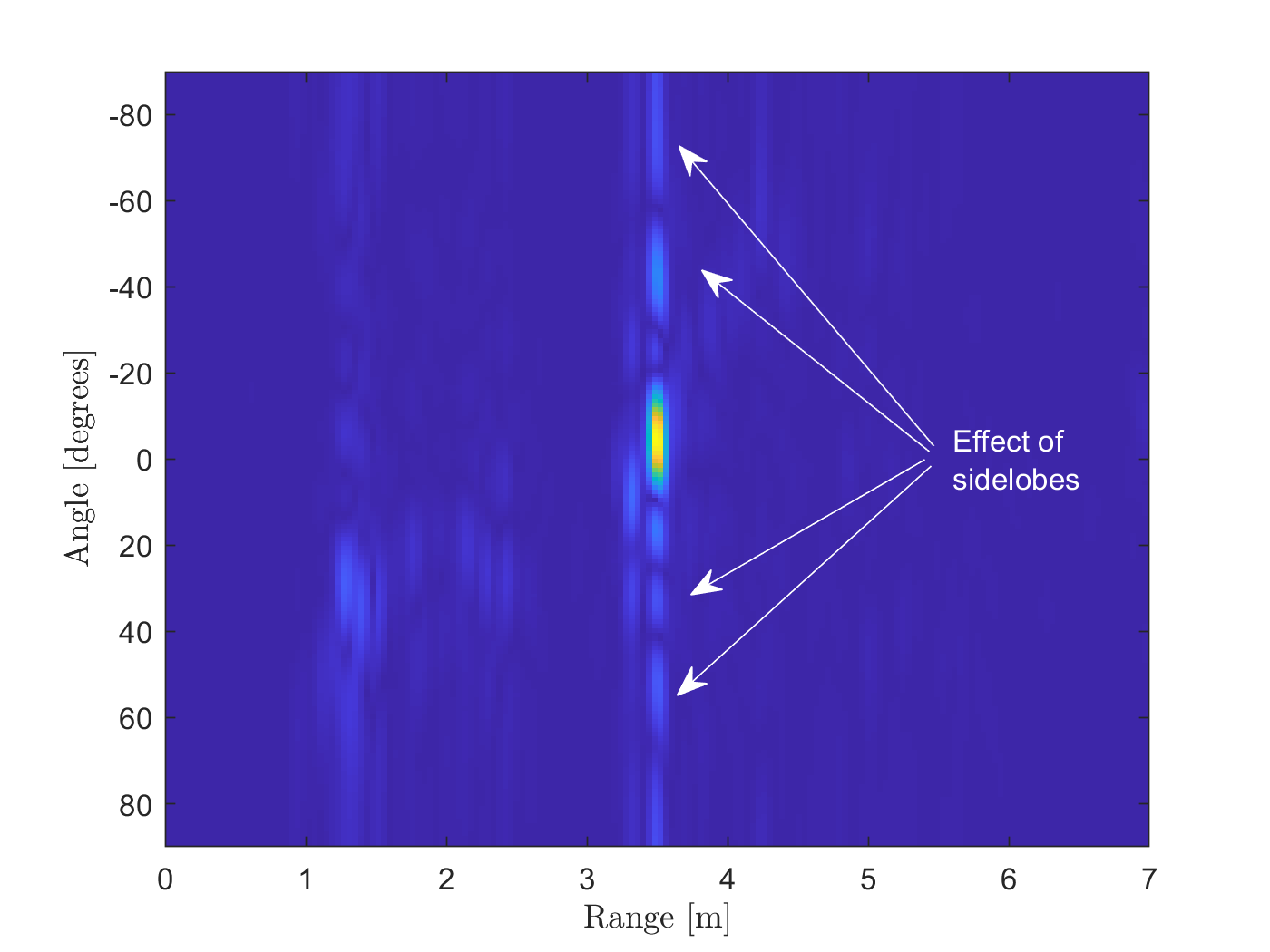}
		\label{fig:Before_clean}} 
	\subfigure[$\mathbf{H}$ matrix after the \ac{GEM} procedure.]
	{\includegraphics[trim= {10 0 35 10}, clip,width=0.4\linewidth]{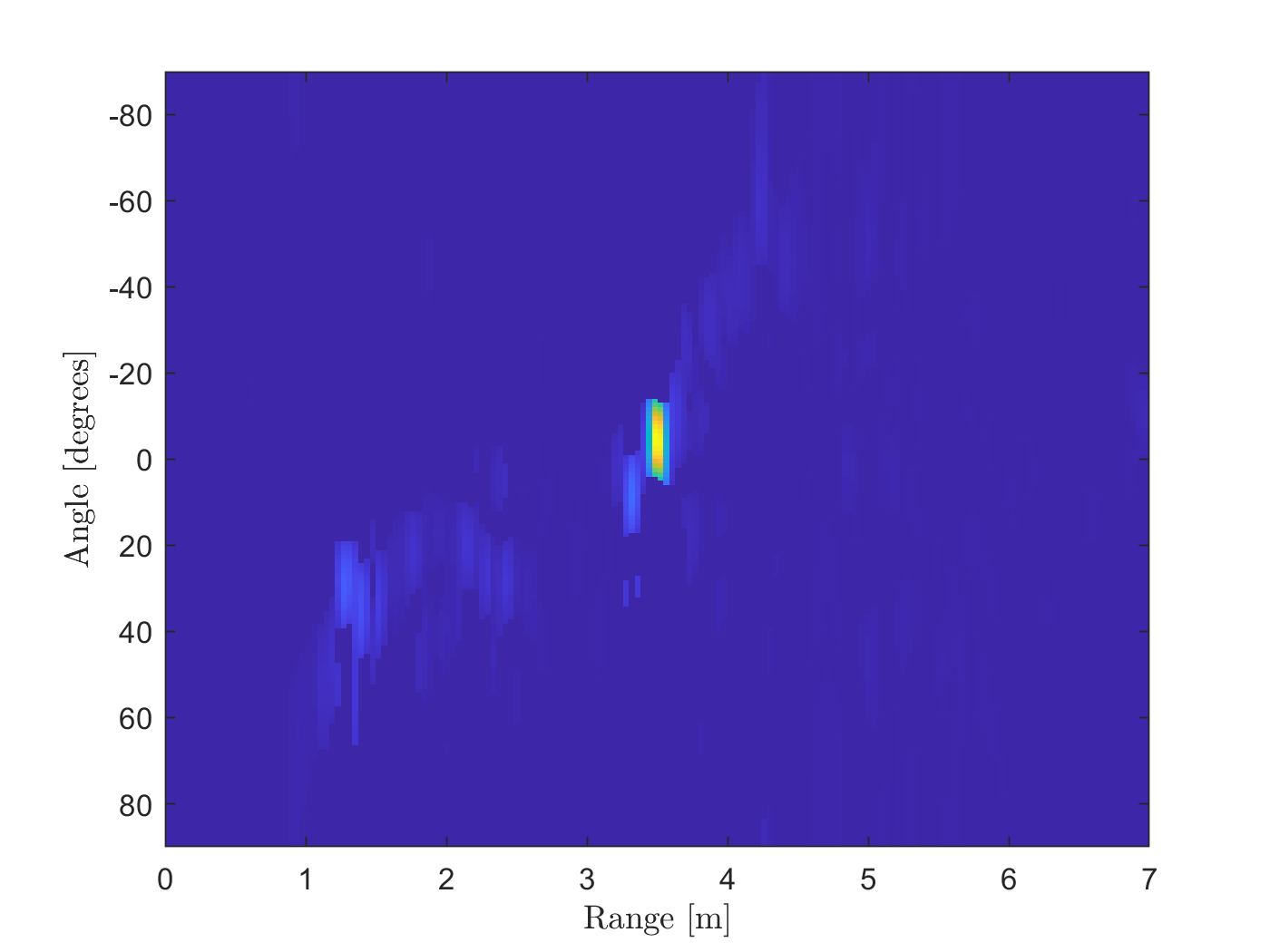}
		\label{fig:After_clean}} 
	\caption{Impact of the \ac{GEM} procedure. Single obstacle located at three meters in front of the radar working at $77\,\mathrm{GHz}$ with $\eta_{\text{CL}}=0.4$.}
	\label{fig:CLEAN} 
\end{figure*}

As shown in Fig.~\ref{fig:Processing_chain}, to mitigate this phenomenon a \acf{GEM} procedure is performed, which operates on each column $\mathbf{h}_m$ of $\mathbf{H}$, with $m=1, 2, \ldots, M$. In particular, for each column vector 
\begin{align}
\mathbf{h}_m=[|h_{1,m}|,|h_{2,m}|,\ldots,\,|h_{n,m}|, \ldots |h_{N,m}|]^T    
\end{align}
of $\mathbf{H}$ (that is, for each distance from the radar), the maximum value is determined as
\begin{align}
{\hat{h}_{m}= \max (\mathbf{h}_m)} 
\end{align}
where $\hat{h}_{m}$ corresponds to the strongest echo detected at the considered distance. Then, we define a threshold
\begin{align}
    \xigem=\eta_{\text{CL}}\cdot \hat{h}_{m},\,\quad 0<\eta_{\text{CL}}\leq1
\end{align}
such that, for each column vector $\mathbf{h}_m$, we  operate as follows
\begin{align}
    |h_{n,m}|=\begin{cases}
       &\!\!\!\!\!\!\!|h_{n,m}| \quad \quad\quad \text{if}\quad |h_{n,m}|\geq \xigem  \\
       & 0 \quad \quad \quad \quad \text{if}\quad |h_{n,m}|<\xigem\,.
    \end{cases}
\end{align}
In other words, by properly defining the parameter $\eta_{\text{CL}}$ 
%a threshold $\xigem$ $\forall\,m=1,\ldots,\,M$ 
that depends on the side lobes level, and hence, on the antenna radiation pattern, it is possible to mitigate the presence of  artifacts.
The rationale behind the \ac{GEM} method proposed is that it is unlikely that multiple obstacles are present exactly at the same distance from the radar, whereas echoes captured by sidelobes appear exactly at the same distance. 
Fig.~\ref{fig:CLEAN} shows an example of the beneficial impact of the \ac{GEM} algorithm on a matrix $\mathbf{H}$ obtained from measurements taken in the $77$ $\mathrm{GHz}$ band. Although in this paper we operate with measurements taken in the [$235-320$] $\mathrm{GHz}$ band, we purposely switched to the $77$ $\mathrm{GHz}$ band to generate Fig.~\ref{fig:CLEAN} to better highlight the ghost artifact phenomenon, which is visually more evident in the millimeter-wave band because of the lower resolution of the radar.

\subsubsection{Noise Masking}
\label{Section:cfar}
Another important impairment affecting the accuracy of measurements is the background noise, either received by the antenna or generated by the radar circuitry itself. Radars usually mitigate its impact by running a \ac{NM} algorithm that eliminates all the signal components that fall below a certain masking threshold. 
%, which define a threshold above which a detected signal is retained because it is considered as coming from a target, as opposed to one generated by spurious sources, which is instead cancelled.
%
In our case, the \ac{NM} algorithm operates as follows. First, it detects the peak of the matrix $\mathbf{H}$, that is,
\begin{align}
h_{\text{max}}=\max(\mathbf{H}) \,.   
\end{align}
Then, similarly to the \ac{GEM} algorithm, it requires to define
the threshold%
\begin{align}
        \xinm=\eta_{\text{CF}}\cdot h_{\text{max}},\,\quad 0<\eta_{\text{CF}}\leq1
\end{align}
where $\eta_{\text{CF}}$  is a properly chosen parameter
that depends on the background noise and it can be based, for instance, on the \ac{CFAR} strategy. Given this threshold, the matrix $\mathbf{H}$ is cleaned of the unwanted noise contribution as follows
\begin{align}
    |h_{n,m}|=\begin{cases}
       &\!\!\!\!\!\!\!|h_{n,m}| \quad \quad\quad \text{if}\quad |h_{n,m}|\geq \xinm  \\
       & 0 \quad \quad \quad \quad \text{if}\quad |h_{n,m}|<\xinm \,.
    \end{cases}
\end{align}

As shown in Fig.~\ref{fig:Processing_chain}, after undergoing the \ac{GEM} and \ac{NM} processings, the response to matrix $\mathbf{H}$ is denoted as $\mathbf{F}_k$ and is referred to as \emph{frame}. The subscript $k$ has been introduced to emphasize that a new frame is generated each time the radar performs a complete scan of the environment, which occurs at time instants $k\, T_{\text{F}}$.
As evident in Fig.~\ref{fig:Processing_chain}, as soon as a new  frame $\mathbf{F}_k$  is generated, it is passed to the subsequent stage, which is in charge of updating the estimates of the mobile user's trajectory and of the environment map. %, which will be discussed in the following. %\subsubsection{SLAM algorithm}. 

\subsection{User's Trajectory Estimation}

As shown in Fig.~\ref{fig:Processing_chain}, the estimation of the user's trajectory is performed by means of a Kalman filter on the basis of the pose estimates $\hat{\mathbf{p}}_k$ obtained by comparing the current frame $\mathbf{F}_k$ with that of the previous time instant, $\mathbf{F}_{k-1}$. In this regard, it is worth emphasizing that $\mathbf{F}_k$ and $\mathbf{F}_{k-1}$ are two-dimensional representations of the scanned area (in polar coordinates), hence they can be meant as ``radar images" of the sensed environment in successive instants. This suggested us to derive the pose estimates by means of algorithms that were initially conceived for image processing.
%Indeed, the specific algorithms we adopted to derive the relative pose estimates represent one of the main contributions of this work, hence will be discussed in detail in Section \ref{Sec.Rel_Pose_Estimates}.

The specific pose estimation algorithms considered in this manuscript, which represent one of the major contributions of our work, deserve detailed descriptions, which are therefore provided in the specially dedicated Section \ref{Sec.Rel_Pose_Estimates}. In any case, whatever pose estimation algorithm is adopted, $\hat{\mathbf{p}}_k$ is derived starting from the estimated relative pose vector 
\begin{equation}\label{eq:usersrelpose}
{\tilde{\mathbf{z}}_k=[\tilde{dx}, \tilde{dy}, \tilde{d\theta}]^T}\,, 
\end{equation}
whose elements are the estimated horizontal shift, vertical shift and rotation of the current pose  with respect to the previous one, all referenced to the radar-based coordinate system,
as well as a quality indicator $q \in [0,1]$ of the estimates.
It should  be emphasized that $\tilde{\mathbf{z}}_k$ refers to the local coordinate system of the mobile user, i.e., the radar point of view, therefore it has to be transformed into the absolute coordinate system by considering the latest available estimate of the mobile user's rotation $\hat{\theta}_{k-1}$, that is,
\begin{equation}\label{eq:op1}
\mathbf{z}_k=\mathbf{U}(\hat{\theta}_{k-1})\,  \tilde{\mathbf{z}}_k
\end{equation}
where
\begin{equation}\label{eq:umat}
\mathbf{U}(\theta)=\left [
\begin{array}{ccc}
    \cos \theta & \sin \theta & 0 \\
    -\sin \theta & \cos \theta & 0 \\
    0 & 0 & 1
\end{array}
\right ]
\end{equation}
denotes the rotation matrix of angle $\theta$.
Given $\mathbf{z}_k$, and the previous absolute pose estimate, denoted by ${\hat{\mathbf{p}}_{k-1}=[\hat{x}_{k-1}, \hat{y}_{k-1}, \hat{\theta}_{k-1} ]^T}$, the current raw absolute pose estimate is 
  \begin{equation}\label{eq:op2}
    \hat{\mathbf{p}}_k= \hat{\mathbf{p}}_{k-1} + \mathbf{z}_k \, .
\end{equation}

% the absolute pose, i.e., the user's trajectory, is derived through a Kalman filter, which accounts for the mobility model defined in \eqref{eq:mobilitymodel}. \red{GP: la sottosezione seguente deve essere rivisitata}.

%\subsubsection{Trajectory Estimation}
The processing described by \eqref{eq:op1}, \eqref{eq:umat} and \eqref{eq:op2}, is represented in Fig.~\ref{fig:Processing_chain} by the block named $\Psi$, which receiving as input $\tilde{\mathbf{z}}_k$ generates as output $\hat{\mathbf{p}}_k$. The latter is then passed to the Kalman filter, which is in charge of the trajectory estimation accounting for the user's mobility model and the quality of the relative pose estimates.
In this regard, we point out that when it comes to tracking algorithms, it is customary to consider a Markovian state-space model to describe the evolution of the state, which is based on the following  second-order kinematic linear model %\cite[Ch. 6, Eq. 6.2.2-10,12]{bar2004estimation} \cite{dardari2015indoor} 
\cite{bar2004estimation,dardari2015indoor} 
% linear model according to the
%
\begin{equation}\label{eq:mobilitymodel}
\textbf{x}_{k+1} = \textbf{A} \,  \textbf{x}_k + \textbf{w}_k
\end{equation}
where $\mathbf{A}$ is the transition matrix and $\mathbf{w}_k \sim \mathcal{N}(\mathbf{0}_{1\times6}; \textbf{Q})$ is the process noise with covariance matrix $\textbf{Q}$ given by \cite{dardari2015indoor}
\begin{align}
&\mathbf{A}=\left [
\begin{array}{ccc}
    \mathbf{I}_{2} & T_{\text{F}} \, \mathbf{I}_{2} & \mathbf{0}_{2} \\
    \mathbf{0}_{2} &  \mathbf{I}_{2} & \mathbf{0}_{2} \\
    \mathbf{0}_{2} &  \mathbf{0}_{2} & \tilde{\mathbf{A}} \\
\end{array}
\right ], \\ &\mathbf{Q}=
\left [
\begin{array}{ccc}
    w_0 \, \frac{T_{\text{F}}^3}{3} \, \mathbf{I}_{2} &  w_0 \, \frac{T_{\text{F}}^2}{2} \, \mathbf{I}_{2}  & \mathbf{0}_{2} \\
     w_0 \, \frac{T_{\text{F}}^2}{2} \, \mathbf{I}_{2}  &  w_0 \, T_{\text{F}} \, \mathbf{I}_{2} & \mathbf{0}_{2} \\
    \mathbf{0}_{2} &  \mathbf{0}_{2} & \tilde{\mathbf{Q}} \\
\end{array}
\right ]
\end{align}
with $\mathbf{I}_{N}$ and $\mathbf{0}_{N}$ being respectively  $N\times N$ identity and zero matrices, $w_0$  being the power spectral density of the linear acceleration noise, and with
% T_{\text{F}}
\begin{align}
    & \tilde{\mathbf{A}}=\left [
\begin{array}{cc} 1 & T_{\text{F}}\\  0& 1 \end{array} \right], &&\tilde{\mathbf{Q}}=\left [
\begin{array}{cc} \omega_\theta \frac{T_{\text{F}}^3}{3} & \omega_\theta \frac{T_{\text{F}}^2}{2}  \\  \omega_\theta \frac{T_{\text{F}}^2}{2} & \omega_\theta T_{\text{F}} \end{array} \right],
\end{align}
where  $w_{\theta}$ is the power spectral density of the angular acceleration noise, which depends on the expected mobility of the user.

The evolution of the absolute state $\mathbf{x}_k$ can be tracked by means of a Kalman filter, fed step-by-step with the current raw absolute pose estimate, $\hat{\mathbf{p}}_k$, using the following observation model 
\begin{equation}\label{eq:globalx}
    \hat{\mathbf{p}}_k=\mathbf{B} \, \mathbf{x}_k + \bm{\nu}_k
\end{equation}
where
\begin{equation}
\mathbf{B}=\left [
\begin{array}{cccccc}
    1 & 0 & 0 & 0 & 0 & 0 \\
    0 & 1 & 0 & 0 & 0 & 0 \\
    0 & 0 &0 & 0 & 1 & 0 
\end{array}
\right ]
\end{equation}
and  $\bm{\nu}_k \sim \mathcal{N}(\mathbf{0}_3; \mathbf{R})$ is the estimation noise modeled as  Gaussian random vector  with covariance matrix \cite{dardari2015indoor}
\begin{align} \label{eq:R}
\mathbf{R}=\text{diag} \left (\sigma_x^2 / q^2, \sigma_y^2/ q^2, \sigma_{\theta}^2/ q^2 \right )
\end{align}
which accounts for the reliability $q$ of the current relative pose estimate, with  $\sigma_x^2= \sigma_y^2$, and $\sigma_{\theta}^2$ being the estimation noise power expected by the specific relative pose estimator.
% of $dx$, $dy$ and $d\theta$, respectively.
At each time instant $k$, the Kalman filter provides an estimate $\hat{\mathbf{x}}_k$ of the state $\mathbf{x}_k$ as well as its covariance matrix.

\subsection{Automatic mapping}
\label{subsec:Mapping}
As for the mapping task, in this paper we consider an \emph{occupancy grid} method to represent the indoor surroundings. Occupancy grid-based methods consist of dividing the scenario into a fine grid of cells and solving the problem of inferring the occupancy probability of each cell based on radar measurements and the estimated trajectory. For each cell, the likelihood of its occupancy must be calculated on the basis of the collected observations. 

% Solving this problem is crucial in various applications, including \ac{SLAM}, as it enhances the radar's knowledge of the surrounding \cite{DurBai:J06},  and hence, the possibility of localizing without the need for a dedicated infrastructure. 

% The map and pose estimations are performed jointly, and high accuracy in map reconstruction allows for better radar position estimation and vice versa \cite{GuiGueDar:J15,guerra2021real}. 
%For autonomous radar navigation, an estimate of the map is also essential for obstacle avoidance \cite{Thr:J02}.
According to \cite{GuiGueDar:J15}, the actual map in the time instant $k$ can be represented as a vector of cells as\footnote{The map might be time varying due to moving obstacles, such humans or mobile furniture (e.g., chairs).} %\textcolor{red}{[Eq. 19 corrisponde a ``map", l'ingresso della funzione ``insertRay"]}
\begin{align}\label{eq:map}
&\mapk \triangleq \left[\mapsk,\, \ldots, \mapik,\, \ldots,\, \mapek \right]^{\mathsf{T}} \in \mathbb{B}^{\Ncell}
\end{align}
where ${\mapik \in \mathbb{B}}$ represents the true occupancy of the $i$-th cell ($\mapik=0$ denotes an empty cell, whereas ${\mapik=1}$ denotes an occupied cell), $\mathbb{B}$ is the Boolean domain, and $\Ncell$ is the total number of considered cells. In the sequel, we consider a stationary map, that is $\mapk= \map$, $\forall k$.

The goal of the mapping algorithm is to infer \eqref{eq:map} by computing the maximum of the \emph{a-posteriori} probability mass function  ({\em belief}) given the history of frame observations and the estimated trajectories (see Fig.~\ref{fig:Processing_chain}). Denoting with $b_k(m_i)$ the belief of occupancy of the $i$-th cell at time instant $k$,  the following steps are performed:
\begin{itemize}
\item \textit{Initialization:}  If no prior map information is available, a possible initialization is $\bs(\mapi)=0.5$, corresponding to a complete uncertainty, $\forall i=1, 2, \ldots, \Ncell$.  %\textcolor{red}{[Questo valore dovrebbe corrispondere a ``map" all'istante iniziale. Ad esempio, in un codice che avevo fatto ho questa linea di codice: map1 = occupancyMap(10,10,mapResolution); con mapResolution = 2. Chiamando getOccupancy(map,[8 5]) ottengo 0.5, quindi dovrebbe essere l'init di default.]}
\item \textit{Scan vector generation:} At the generic time instant $k$, the cleaned Angle-Range matrix $\mathbf{F}_{k}$, obtained as output of the \ac{GEM} and \ac{NM} processing,  is passed to the mapping algorithm (the Mapping block in Fig.~\ref{fig:Processing_chain}),
whose first task is to generate a 1-by-$N$ vector of ranges $\mathbf{v}_k$ similar to the one provided by a \ac{Lidar} scan. Specifically, $\mathbf{v}_k$ contains only one range value of the current \emph{frame} $\mathbf{F}_k$  for each considered steering angle, i.e., ${\mathbf{v}_k=\left [v_1^{(k)}, v_2^{(k)}, \ldots, v_N^{(k)} \right ]}$, with $v_n^{(k)}$ being the range of the object (if any) seen at the angle $\phi_n$ by the radar. %(this entails that a single range value is saved in the vector for each row of $\mathbf{F}_k$). 
This result is achieved by comparing each row of the current frame, i.e., ${\mathbf{f}_n=[|h_{n,1}|,|h_{n,2}|, \cdots |h_{n,M}|]}$  of $\mathbf{F}_k$, with a suitable threshold ${0<\eta_{\text{SV}}\leq1}$. The distance corresponding to the first element which exceeds $\eta_{\text{SV}} \cdot f_{\text{max}}$, with  $f_{\text{max}}$ being the maximum value in $\mathbf{f}_n$, is saved in  $\mathbf{v}_k$. % if that condition is not satisfied by any element of the row, being $f_{\text{max}}$  the maximum value in $\mathbf{f}_n$. 

Similarly, the angles $\phi_n$ are collected into the angle vector $\bm{\phi}=[\phi_1, \phi_2, \ldots, \phi_N]$. 
The final scan vector at time instant $k$ is given by $\mathbf{s}_k=\left[ \mathbf{v}_k^T,\, \bm{\phi}^T\right]$. 
% The position of each element of $\mathbf{F}_{k}$ different from zero corresponds to the polar coordinates of a potential obstacle relative to the current user's position and rotation.  By considering the estimated radar state $\hat{\mathbf{x}}_k$, the relative coordinates of each detected obstacle are translated into a set of absolute cartesian coordinates and   collected in the vector  $\mathbf{s}_k$.

\item \textit{Log-Odd Update:} Starting from $\mathbf{s}_k$, the beliefs are updated following a classic occupancy grid algorithm \cite{Thr:J98}.  To avoid numerical instability, a practical solution is to map the belief into the log-odd quantity  
\begin{align}
\logoddk\left( \mapi \right) \triangleq \log\left(\frac{\bk(\mapi)}{1-\bk(\mapi)} \right),\,\forall\, i=1, 2, \ldots, \Ncell
\end{align}
and consider the  log-odd update given by \cite{Thr:J98} %\textcolor{red}{[corrisponde a ``insertRay"]}
 \begin{align}\label{eq:logoddupdate} \logoddk\left(\mapi\right)\!=&\!\log\left(\frac{p \left(\mathbf{s}_k \lvert \mapi =1   \right) }{{p \left( \mathbf{s}_k \lvert \mapi=0  \right)}} \right) +\logoddkprev\left(\mapi \right)\nonumber \\ & \forall\, i=1, \ldots, \Ncell
\end{align}
where $p \left(\mathbf{s}_k \lvert \mapi =1   \right)$ ($p \left(\mathbf{s}_k \lvert \mapi =0   \right)$) is the likelihood function considering the current scan $\mathbf{s}_k$ given the presence of an occupied (free) cell in $m_i$. In our case, $p \left(\mathbf{s}_k \lvert \mapi =1   \right)$ can take two values, that is $0.9$ when the polar coordinates of the $i$th cell are present inside the scan vector $\mathbf{s}_k$,  and $0.1$ otherwise.  Finally, the likelihood of having an empty cell is simply given by  
\begin{align}
    p \left(\mathbf{s}_k \lvert \mapi =0   \right)=1- p \left(\mathbf{s}_k \lvert \mapi =1\right)\,.
\end{align}
\end{itemize}

In \eqref{eq:logoddupdate} we assume that each cell is independent from all the others (including adjacent cells) as for laser observations. Nevertheless, inter-cell correlations can be considered in the observation model to further refine the mapping process \cite{GueEtAl:C18}.

\section{Relative Pose Estimation}

\label{Sec.Rel_Pose_Estimates}
In this section, we provide the details of the algorithms we considered for the estimation of the relative pose, which refer to the Pose Estimation block in Fig.~\ref{fig:Processing_chain}. First, we introduce the  \emph{Laser Scan Matching} algorithm \cite{HeKoRaAn:C16}, commonly adopted in the case of \ac{Lidar}-based systems, which is considered here as a benchmark.
Then, inspired by the possibility of employing algorithms typically used for image processing, we propose a modified version of the \emph{Fourier-Mellin} algorithm, specifically adapted to operate with signals provided by a THz radar. Finally, we propose a simplified version of the \emph{Fourier-Mellin}-based algorithm that, having lower computational complexity, is better suited for use in portable devices implementing \ac{R-SLAM}. 

% In this section, inspired by the possibility to employ algorithms typically used for image processing and laser-\ac{SLAM}, we first discuss and tailor to the \ac{THz} scenario two state-of-the-art approaches for the relative pose estimation, namely the  \emph{Laser Scan Matching} \cite{HeKoRaAn:C16} and the \emph{Fourier-Mellin} algorithms. Then, we propose a simplified \emph{Fourier-Mellin}-inspired method that better fits for implementation in low-complex devices (e.g., \emph{personal radars} and autonomous agents) for \ac{R-SLAM} applications.

\paragraph{Laser Scan Matching Algorithm}
This methodology was conceived for \ac{Lidar}-based systems and it is here considered as a benchmark \cite{HeKoRaAn:C16}. When fed with laser scans, it has been proved to achieve real-time loop closure and $5\,$cm resolution \cite{HeKoRaAn:C16}. %Solid implementations of this algorithms exist and have also been included in the MATLAB Lidar Toolbox.  

According to this algorithm, the user's relative pose in the current instant $k$ in \eqref{eq:usersrelpose}  is derived starting from the current and the previous scan vectors $\mathbf{s}_k$ and $\mathbf{s}_{k-1}$, which are compared to estimate the corresponding translation increments $(\tilde{dx}, \tilde{dy})$ and rotation increment $\tilde{d\theta}$ using a grid-based search.\footnote{The scan vectors $\mathbf{s}_k$ and $\mathbf{s}_{k-1}$ on which the \emph{Laser Scan Matching} algorithm operates are derived as explained in Sec.\ref{subsec:Mapping}}
%
% Starting from the \emph{frames} $\mathbf{F}_k$ provided by the pre-processing stage of Fig.~\ref{fig:Processing_chain}, the first step is to %distil the information content of the current \emph{frame} $\mathbf{F}_k$ into 
% form a 1-by-$N$ vector of ranges, namely $\mathbf{v}_k$, similar to the one provided by a \ac{Lidar} scan, where only one range value of the current \emph{frame} $\mathbf{F}_k$  is returned for each considered steering angle, i.e., ${\mathbf{v}_k=[v_1^{(k)}, v_2^{(k)}, \ldots, v_N^{(k)}]}$, with $v_n^{(k)}$ corresponding to the angle $\phi_n$. %(this entails that a single range value is saved in the vector for each row of $\mathbf{F}_k$). 
% This result is achieved by comparing each row of the current frame, i.e., $\mathbf{f}_n=[|h_{n,1}|,|h_{n,2}|, \cdots |h_{n,M}|]$  of $\mathbf{F}_k$, with a suitable threshold ${0<\eta_{\text{SV}}\leq1}$. The distance corresponding to the first element which exceeds $\eta_{\text{SV}} \cdot f_{\text{max}}$, with  $f_{\text{max}}$ being the maximum value in $\mathbf{f}_n$, is saved in  $\mathbf{v}_k$. % if that condition is not satisfied by any element of the row, being $f_{\text{max}}$  the maximum value in $\mathbf{f}_n$. 
%
% Similarly, the angles $\phi_n$ are collected into the angle vector $\bm{\phi}=[\phi_1, \phi_2, \ldots, \phi_N]$. 
% The final scan vector at time instant $k$ is given by $\mathbf{s}_k=\left[ \mathbf{v}_k^T,\, \bm{\phi}^T\right]$. \textcolor{blue}{AG: Forse possiamo chiamarlo $\tilde{\mathbf{s}}_k$ per distinguerlo da quello usato nel mapping con coordinate cartesiane, o viceversa}
\begin{figure*}
\centering
\input{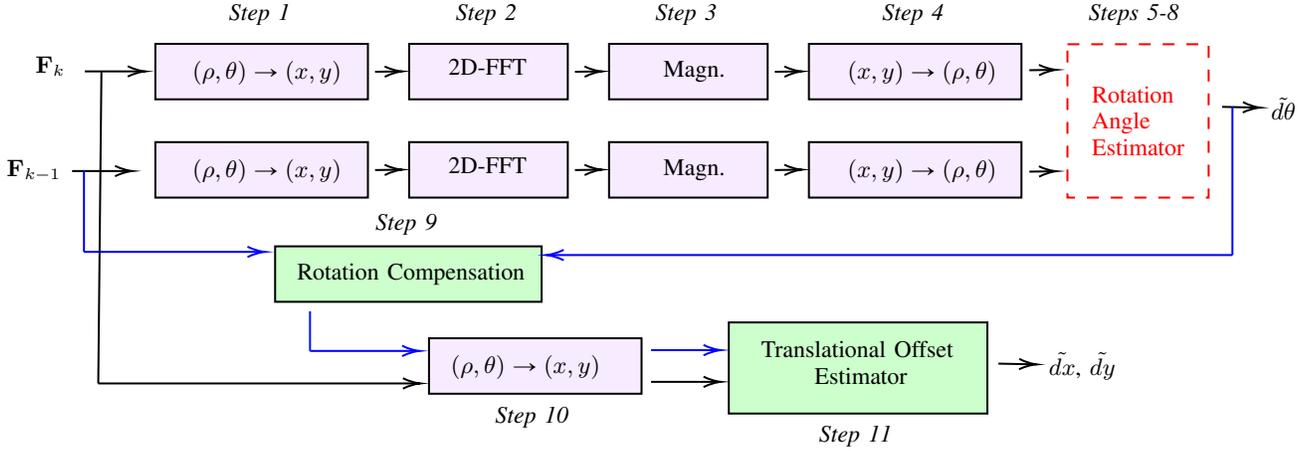}
    \caption{Relative pose estimation implemented using the Fourier-Mellin-based algorithm. The output is the  relative pose ${\tilde{\mathbf{z}}_k=[\tilde{dx}, \tilde{dy}, \tilde{d\theta}]^T}$. The details of the``Rotation Angle Estimator" are depicted in Fig.~\ref{fig:Approx_Fourier_Mellin_Processing}.}  \label{fig:Fourier_Mellin_Processing}
\end{figure*}
% The user's relative pose in the current instant $k$ is then derived through the {\em Laser Scan Matching} algorithm \cite{HeKoRaAn:C16}, which receives as input the current and the previous scan vectors $\mathbf{s}_k$ and $\mathbf{s}_{k-1}$, and estimates the corresponding translation increments $(\tilde{dx}, \tilde{dy})$ and rotation increment $\tilde{d\theta}$ using a grid-based search. 
%
Specifically, the {\em Laser Scan Matching} algorithm converts $\mathbf{v}_k$, $\mathbf{v}_{k-1}$ and $\bm{\phi}$ into probabilistic grids and finds the pose between the two scans by correlating their grids \cite{HeKoRaAn:C16}. The interested reader is referred to \cite{Clausen2003BranchAB} for additional details on the strategy adopted by the algorithm to speed up the computation.  

% The algorithm uses
% a branch-and-bound strategy to speed up computation over large discretized search windows \cite{Clausen2003BranchAB}.  

\paragraph{Fourier-Mellin-based Algorithm}
Aiming to obtain better performance than that offered by the {\em Laser Scan Matching} algorithm, in this paper we propose to estimate the relative pose between two consecutive frames  $\mathbf{F}_k$ and $\mathbf{F}_{k-1}$ by means of the Fourier-Mellin algorithm \cite{QinDefDec:J94,RedCha:J96}, which is an FFT-based method used to register\footnote{Image registration is an image processing technique used to align multiple scenes into a single integrated image, compensating rotations, translations and different scaling.} two different images searching for the optimal match in the frequency domain. In particular, the objective of the  algorithm is to decouple the translation and rotation effects in order to facilitate their estimation. 

This algorithm, which was designed to operate on Cartesian images, is here applied to consecutive \emph{frames} in polar coordinates, which in fact can be interpreted as \quot{radio images} of the environment. For the sake of clarity, in the following we will highlight the dependence of the elements of  $\mathbf{F}_k$ on the distance $\rho$ and the angle $\theta$ by treating the frame as a two-dimensional function  $\mathbf{F}_k(\theta, \rho)$. 
 It is worth pointing out, in this regard, that $\rho$ and $\theta$ are here meant as continuous variables, in contrast with the corresponding  discrete variables $d_m$ and $\phi_n$ introduced in Sec.~\ref{Sec:RSLAM}. This choice is aimed at simplifying the notation in the following analysis, the discrete version of which was implemented in our test bed.

The algorithm we propose consists of the main steps shown in Figs.~\ref{fig:Fourier_Mellin_Processing}-\ref{fig:Approx_Fourier_Mellin_Processing}, which are described hereafter.

\begin{itemize}
    \item {\em Step 1}. The current and previous frames $\mathbf{F}_k(\theta, \rho)$ and $\mathbf{F}_{k-1}(\theta, \rho)$, which clearly refer to a polar coordinate system, are converted into the corresponding images  $\mathbf{C}_k(x,y)$ and $\mathbf{C}_{k-1}(x,y)$ in Cartesian coordinates.
    \item {\em Step 2}. Given $\mathbf{C}_k(x,y)$ and $\mathbf{C}_{k-1}(x,y)$, the corresponding 2D Fourier transforms $\mathbfcal{F}_k(\xi, \eta)$ and $\mathbfcal{F}_{k-1}(\xi, \eta)$ are computed. Assuming that $\mathbf{C}_k(x,y)$ is a perfectly (i.e., not affected by noise and artifacts) rotated and translated replica of  $\mathbf{C}_{k-1}(x,y)$, that is
    \begin{align}
    \mathbf{C}_{k}(x,y)=\mathbf{C}_{k-1}&(x\cos(d\theta)+y\sin(d\theta)-dx, -x\sin(d\theta) \nonumber\\
    &+y\cos(d\theta)-dy) \,
    \end{align}
    %
    %\textcolor{red}{AG: un'alternativa pi\'u compatta potrebbe essere:
    %
    %
    %\begin{align}
    %&\mathbf{C}_{k}(x,y)=\mathbf{R}\left(d\theta\right)\mathbf{C}_{k-1}(x-dx,y-dy) 
    %\end{align}
    %where $\mathbf{R}\left(d\theta\right)$ is a rotation matrix defined as $\mathbf{R}\left(d\theta\right)=[\cos(d\theta),\, \sin(d\theta); \cos(d\theta),\, -\sin(d\theta)]$ OPPURE $\mathbf{R}\left(d\theta\right)=\left[\mathbf{U}(d\theta)\right]_{1:2,1:2}$. (la seconda opzione \'e valida se rimane l'eq. \eqref{eq:umat}.)
    %}
   where $(dx,dy)$ are the translational offsets and $d\theta$ is the rotation angle from instant $k-1$ to instant $k$, the relation between the corresponding  2D  Fourier transforms is:
    \begin{align}
    \mathbfcal{F}_k(\xi, \eta)=e^{-j2\pi(\xi dx+\eta dy)}\mathbfcal{F}_{k-1}&(\xi\cos(d\theta)+\eta\sin(d\theta), \nonumber \\ &-\xi\sin(d\theta)+\eta\cos(d\theta)).
    \label{eq:transforms_relation}
    \end{align}
 %   \textcolor{red}{
%\begin{align}
%    &\mathbfcal{F}_k(\xi, \eta)=e^{-j2\pi(\xi dx+\eta dy)}\, \left[\mathbf{R}\left( d\theta\right)\mathbfcal{F}_{k-1}(\xi,\eta)\right].
%    \label{eq:transforms_relation}
%    \end{align}
%    }
    \item {\em Step 3}. Given \eqref{eq:transforms_relation}, the relation between the magnitudes of $\mathbfcal{F}_k(\xi, \eta)$ and $\mathbfcal{F}_{k-1}(\xi, \eta)$ is
    \begin{align}
    \mathbf{M}_k(\xi, \eta)=\mathbf{M}_{k-1}(&\xi\cos(d\theta)+\eta\sin(d\theta), \nonumber \\ &-\xi\sin(d\theta)+\eta\cos(d\theta)).
    \label{eq:magnitude_relation}
    \end{align}
%\textcolor{red}{
%    \begin{align}
%    &\mathbf{M}_k(\xi, \eta)= \mathbf{R}(d\theta)\mathbf{M}_{k-1}(\xi,\eta).
%    \label{eq:magnitude_relation}
%    \end{align}
%    }
    
    From \eqref{eq:magnitude_relation} one can argue that the magnitude is translation invariant, as it does not depend on $(dx,dy)$. It turns out, therefore, that possible differences between $\mathbf{M}_k(\xi, \eta)$ and $\mathbf{M}_{k-1}(\xi, \eta)$ depend  on the  rotation $d\theta$ only.
    \item {\em Step 4}. By expressing \eqref{eq:magnitude_relation} in polar coordinates, it immediately results
    \begin{equation}
    \mathbf{M}_k(\theta, \rho)=\mathbf{M}_{k-1}( \theta-d\theta, \rho),
    \label{eq:magnitude_relation_polar}
    \end{equation}
    that represents a convenient formulation for the derivation of the rotation angle $d\theta$. Actually, converting the magnitudes from Cartesian to polar coordinates makes it possible to represent rotations as  translations in the angular domain, as evident in \eqref{eq:magnitude_relation_polar}, thus allowing to exploit the translation property of the Fourier transform, as explained in Step 5.
    \item {\em Step 5.} By denoting $\mathbfcal{M}_k(\mu, \nu)$ and $\mathbfcal{M}_{k-1}(\mu, \nu)$ the 2D Fourier transforms of  $\mathbf{M}_k(\theta, \rho)$ and $\mathbf{M}_{k-1}(\theta, \rho)$, respectively, it results 
    \begin{equation}
        \mathbfcal{M}_k(\mu, \nu)=\mathbfcal{M}_{k-1}(\mu, \nu)\, e^{-j 2 \pi \nu d\theta}
        \label{eq:magnitude_trasnform}
    \end{equation}
    that features the nice property of having the rotation angle $d\theta$ included only in the exponential function, which can be easily isolated, as shown in Step 6.
    \item {\em Step 6.} The cross-power spectrum $\mathsf{CPS}_k(\mu, \nu)$ of $\mathbf{M}_k(\theta, \rho)$ and $\mathbf{M}_{k-1}(\theta, \rho)$ is defined as
    \begin{equation}
     \mathsf{CPS}_k(\mu, \nu)=\frac{\mathbfcal{M}_k(\mu, \nu)\, \mathbfcal{M}_{k-1}^*(\mu, \nu)}{|\mathbfcal{M}_k(\mu, \nu) \, \mathbfcal{M}_{k-1}^*(\mu, \nu)|} \, .
    \end{equation}
    %where $(\cdot )^*$ 
    %$\mathbfcal{M}_{k-1}^*(\mu, \nu)$ is 
    %denotes the  complex conjugate.
    %of $\mathbfcal{M}_{k-1}(\mu, \nu)$. 
\begin{figure*}
\centering
\tikzset{every picture/.style={line width=0.75pt,font=\small}} %set default line width to 0.75pt        

\begin{tikzpicture}[x=0.6pt,y=0.6pt,yscale=-1,xscale=1]
%uncomment if require: \path (0,310); %set diagram left start at 0, and has height of 310

%Shape: Rectangle [id:dp38331830319223825] 
\draw  [fill={rgb, 255:red, 247; green, 235; blue, 255 }  ,fill opacity=1 ] (93,55.5) -- (193,55.5) -- (193,90.5) -- (93,90.5) -- cycle ;
%Shape: Rectangle [id:dp23167463729885696] 
\draw  [fill={rgb, 255:red, 247; green, 235; blue, 255 }  ,fill opacity=1 ] (93,100.5) -- (193,100.5) -- (193,135.5) -- (93,135.5) -- cycle ;
%Shape: Rectangle [id:dp3389305696846927] 
\draw  [fill={rgb, 255:red, 0; green, 255; blue, 0 }  ,fill opacity=0.2 ] (223,51) -- (330,51) -- (330,140) -- (223,140) -- cycle ;
%Shape: Rectangle [id:dp9385230426990023] 
\draw  [fill={rgb, 255:red, 247; green, 235; blue, 255 }  ,fill opacity=1 ] (365,78) -- (465,78) -- (465,113) -- (365,113) -- cycle ;
%Shape: Rectangle [id:dp7664959845090602] 
\draw  [fill={rgb, 255:red, 0; green, 255; blue, 0 }  ,fill opacity=0.2 ] (501,78) -- (601,78) -- (601,113) -- (501,113) -- cycle ;
%Straight Lines [id:da39207377795337894] 
\draw    (470,95.25) -- (492,95.71) ;
\draw [shift={(494,95.75)}, rotate = 181.19] [color={rgb, 255:red, 0; green, 0; blue, 0 }  ][line width=0.75]    (10.93,-3.29) .. controls (6.95,-1.4) and (3.31,-0.3) .. (0,0) .. controls (3.31,0.3) and (6.95,1.4) .. (10.93,3.29)   ;
%Straight Lines [id:da40201902037532] 
\draw    (338,95.25) -- (360,95.71) ;
\draw [shift={(362,95.75)}, rotate = 181.19] [color={rgb, 255:red, 0; green, 0; blue, 0 }  ][line width=0.75]    (10.93,-3.29) .. controls (6.95,-1.4) and (3.31,-0.3) .. (0,0) .. controls (3.31,0.3) and (6.95,1.4) .. (10.93,3.29)   ;
%Straight Lines [id:da018059119749245456] 
\draw    (194.5,70.75) -- (216.5,71.21) ;
\draw [shift={(218.5,71.25)}, rotate = 181.19] [color={rgb, 255:red, 0; green, 0; blue, 0 }  ][line width=0.75]    (10.93,-3.29) .. controls (6.95,-1.4) and (3.31,-0.3) .. (0,0) .. controls (3.31,0.3) and (6.95,1.4) .. (10.93,3.29)   ;
%Straight Lines [id:da8844665988921527] 
\draw    (194.5,119.75) -- (216.5,120.21) ;
\draw [shift={(218.5,120.25)}, rotate = 181.19] [color={rgb, 255:red, 0; green, 0; blue, 0 }  ][line width=0.75]    (10.93,-3.29) .. controls (6.95,-1.4) and (3.31,-0.3) .. (0,0) .. controls (3.31,0.3) and (6.95,1.4) .. (10.93,3.29)   ;

%Straight Lines [id:da21220349528678817] 
\draw    (611,95.25) -- (655,95.49) ;
\draw [shift={(657,95.5)}, rotate = 180.31] [color={rgb, 255:red, 0; green, 0; blue, 0 }  ][line width=0.75]    (10.93,-3.29) .. controls (6.95,-1.4) and (3.31,-0.3) .. (0,0) .. controls (3.31,0.3) and (6.95,1.4) .. (10.93,3.29)   ;
%Straight Lines [id:da47178380841557255] 
\draw    (30,71) -- (66,71) ;
\draw [shift={(68,71)}, rotate = 180] [color={rgb, 255:red, 0; green, 0; blue, 0 }  ][line width=0.75]    (10.93,-3.29) .. controls (6.95,-1.4) and (3.31,-0.3) .. (0,0) .. controls (3.31,0.3) and (6.95,1.4) .. (10.93,3.29)   ;
%Straight Lines [id:da7387409769267257] 
\draw    (31,114) -- (67,114) ;
\draw [shift={(69,114)}, rotate = 180] [color={rgb, 255:red, 0; green, 0; blue, 0 }  ][line width=0.75]    (10.93,-3.29) .. controls (6.95,-1.4) and (3.31,-0.3) .. (0,0) .. controls (3.31,0.3) and (6.95,1.4) .. (10.93,3.29)   ;
%Shape: Rectangle [id:dp032270465757163525] 
\draw  [fill={rgb, 255:red, 0; green, 255; blue, 0 }  ,fill opacity=0.2] (118,183) -- (286,183) -- (286,218) -- (118,218) -- cycle ;
%Shape: Rectangle [id:dp4434141229775037] 
\draw  [fill={rgb, 255:red, 0; green, 255; blue, 0 }  ,fill opacity=0.2 ] (404,243) -- (567,243) -- (567,300.5) -- (404,300.5) -- cycle ;
%Straight Lines [id:da41745561250830376] 
\draw [color={rgb, 255:red, 0; green, 0; blue, 255 }  ,draw opacity=1 ]   (140,261) -- (207,261) ;
\draw [shift={(209,261)}, rotate = 180] [color={rgb, 255:red, 0; green, 0; blue, 255 }  ,draw opacity=1 ][line width=0.75]    (10.93,-3.29) .. controls (6.95,-1.4) and (3.31,-0.3) .. (0,0) .. controls (3.31,0.3) and (6.95,1.4) .. (10.93,3.29)   ;
%Shape: Rectangle [id:dp1866822481651227] 
\draw  [fill={rgb, 255:red, 247; green, 235; blue, 255 }  ,fill opacity=1 ] (215,253) -- (349,253) -- (349,288) -- (215,288) -- cycle ;
%Straight Lines [id:da42077441741321686] 
\draw    (574,269) -- (596,269.46) ;
\draw [shift={(598,269.5)}, rotate = 181.19] [color={rgb, 255:red, 0; green, 0; blue, 0 }  ][line width=0.75]    (10.93,-3.29) .. controls (6.95,-1.4) and (3.31,-0.3) .. (0,0) .. controls (3.31,0.3) and (6.95,1.4) .. (10.93,3.29)   ;
%Shape: Rectangle [id:dp46594242167707467] 
\draw  [color={rgb, 255:red, 245; green, 35; blue, 35 }  ,draw opacity=1 ][dash pattern={on 4.5pt off 4.5pt}][line width=0.75]  (82.5,7) -- (617.5,7) -- (617.5,155.88) -- (82.5,155.88) -- cycle ;
%Straight Lines [id:da15296775759988335] 
\draw    (40,70.5) -- (41,280) ;
%Straight Lines [id:da9861843134358594] 
\draw    (41,280) -- (207,280.49) ;
\draw [shift={(209,280.5)}, rotate = 180.17] [color={rgb, 255:red, 0; green, 0; blue, 0 }  ][line width=0.75]    (10.93,-3.29) .. controls (6.95,-1.4) and (3.31,-0.3) .. (0,0) .. controls (3.31,0.3) and (6.95,1.4) .. (10.93,3.29)   ;
%Straight Lines [id:da4218315074720702] 
\draw [color={rgb, 255:red, 0; green, 0; blue, 255 }  ,draw opacity=1 ]   (50,114) -- (51,200.5) ;
%Straight Lines [id:da8916678756777106] 
\draw [color={rgb, 255:red, 0; green, 0; blue, 255 }  ,draw opacity=1 ]   (51,200.5) -- (111,200.5) ;
\draw [shift={(113,200.5)}, rotate = 180] [color={rgb, 255:red, 0; green, 0; blue, 255 }  ,draw opacity=1 ][line width=0.75]    (10.93,-3.29) .. controls (6.95,-1.4) and (3.31,-0.3) .. (0,0) .. controls (3.31,0.3) and (6.95,1.4) .. (10.93,3.29)   ;
%Straight Lines [id:da26006080999948544] 
\draw [color={rgb, 255:red, 0; green, 0; blue, 255 }  ,draw opacity=1 ]   (630,99.5) -- (631,199.5) ;
%Straight Lines [id:da8799719257726111] 
\draw [color={rgb, 255:red, 0; green, 0; blue, 255 }  ,draw opacity=1 ]   (631,199.5) -- (291,200.49) ;
\draw [shift={(289,200.5)}, rotate = 359.83] [color={rgb, 255:red, 0; green, 0; blue, 255 }  ,draw opacity=1 ][line width=0.75]    (10.93,-3.29) .. controls (6.95,-1.4) and (3.31,-0.3) .. (0,0) .. controls (3.31,0.3) and (6.95,1.4) .. (10.93,3.29)   ;
%Straight Lines [id:da4584351081078857] 
\draw    (355,280.5) -- (397,280.5) ;
\draw [shift={(399,280.5)}, rotate = 180] [color={rgb, 255:red, 0; green, 0; blue, 0 }  ][line width=0.75]    (10.93,-3.29) .. controls (6.95,-1.4) and (3.31,-0.3) .. (0,0) .. controls (3.31,0.3) and (6.95,1.4) .. (10.93,3.29)   ;
%Straight Lines [id:da4440626596802242] 
\draw [color={rgb, 255:red, 0; green, 22; blue, 255 }  ,draw opacity=1 ]   (140,222) -- (140,261) ;
%Straight Lines [id:da14257357296453255] 
\draw [color={rgb, 255:red, 0; green, 0; blue, 255 }  ,draw opacity=1 ]   (355,260.5) -- (397,260.5) ;
\draw [shift={(399,260.5)}, rotate = 180] [color={rgb, 255:red, 0; green, 0; blue, 255 }  ,draw opacity=1 ][line width=0.75]    (10.93,-3.29) .. controls (6.95,-1.4) and (3.31,-0.3) .. (0,0) .. controls (3.31,0.3) and (6.95,1.4) .. (10.93,3.29)   ;

% Text Node
\draw (116,63) node [anchor=north west][inner sep=0.75pt]   [align=left] {2D-FFT};
% Text Node
\draw (432,13) node [anchor=north west][inner sep=0.75pt]  [color={rgb, 255:red, 255; green, 0; blue, 0 }  ,opacity=1 ] [align=left] {Rotation Angle Estimator};
% Text Node
\draw (636,69) node [anchor=north west][inner sep=0.75pt]   [align=left] {$\tilde{d\theta}$};
% Text Node
\draw (604,260) node [anchor=north west][inner sep=0.75pt]   [align=left] {$\tilde{dx}$,$\tilde{dy}$};
% Text Node
\draw (6,64) node [anchor=north west][inner sep=0.75pt]   [align=left] {$\mathbf{F}_{k}$};
% Text Node
\draw (3,107) node [anchor=north west][inner sep=0.75pt]   [align=left] {\!\!\!\!\! \!\!\!\!\! $\mathbf{F}_{k-1}$};
% Text Node
\draw (322.85,95.5) node [anchor=east] [inner sep=0.75pt]   [align=left] {\begin{minipage}[lt]{64.5pt}\setlength\topsep{0pt}
\begin{center}
\quad \, Cross-Power \\\quad \, Spectrum
\end{center}

\end{minipage}};
% Text Node
\draw (415,252.75) node [anchor=north west][inner sep=0.75pt]   [align=left] {\begin{minipage}[lt]{90.6pt}\setlength\topsep{0pt}
\begin{center}
\!\!\!\!\!\!\!\! Translational Offset\\
\!\!\!\!\!\!\!\!Estimator
\end{center}

\end{minipage}};
% Text Node
\draw (511,87) node [anchor=north west][inner sep=0.75pt]   [align=left] {Max-search};
% Text Node
\draw (122.5,192) node [anchor=north west][inner sep=0.75pt]   [align=left] {\,\,\,Rotation Compensation};
% Text Node
\draw (116,109.5) node [anchor=north west][inner sep=0.75pt]   [align=left] {2D-FFT};
% Text Node
\draw (386,87) node [anchor=north west][inner sep=0.75pt]   [align=left] {2D-IFFT};
% Text Node
\draw (235,262) node [anchor=north west][inner sep=0.75pt]   [align=left] {$(\rho, \theta) \rightarrow (x,y)$};
% Text Node
\draw (120,25) node [anchor=north west][inner sep=0.75pt]   [align=left] {\textit{Step 5}};
% Text Node
\draw (252,25) node [anchor=north west][inner sep=0.75pt]   [align=left] {\textit{Step 6}};
% Text Node
\draw (388,54.5) node [anchor=north west][inner sep=0.75pt]   [align=left] {\textit{Step 7}};
% Text Node
\draw (526,54.5) node [anchor=north west][inner sep=0.75pt]   [align=left] {\textit{Step 8}};

\end{tikzpicture}
    \caption{Relative pose estimation implemented using the simplified Fourier-Mellin algorithm. The output is the  relative pose ${\tilde{\mathbf{z}}_k=[\tilde{dx}, \tilde{dy}, \tilde{d\theta}]^T}$.}  \label{fig:Approx_Fourier_Mellin_Processing}
\end{figure*}
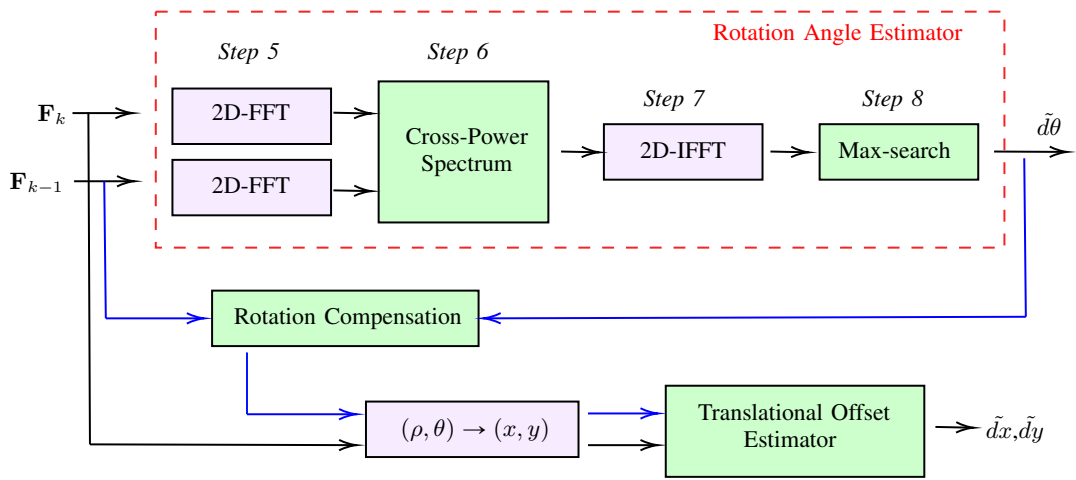
    
    Given \eqref{eq:magnitude_trasnform}, it immediately follows 
        \begin{equation}
      \mathsf{CPS}_k(\mu, \nu)=e^{-j 2 \pi \nu d\theta}.
       \label{eq:CPS_Fourier_Mellin}
    \end{equation}
    \item {\em Step 7}. Taking the inverse Fourier transform of \eqref{eq:CPS_Fourier_Mellin} yields a Dirac $\delta$-function centered at $d\theta$. 
    \item {\em Step 8}. Finding the location where the maximum of the inverse Fourier transform occurs, allows to derive an estimation  $\tilde{d\theta}$ of the rotation angle $d\theta$. In case of less than perfect images, the peak amplitude can be used as a  quality indicator $q$ of the relative pose estimation to be used in \eqref{eq:R}.
    \item {\em Step 9}. Given $\tilde{d\theta}$, it is now possible to apply such rotation to $\mathbf{F}_{k-1}(\theta, \rho)$, thus obtaining  $\overset{\curvearrowright}{\mathbf{F}}_{k-1}(\theta, \rho)$, which (ideally) is aligned in the angle domain with $\mathbf{F}_{k}(\theta, \rho)$.   
    \item {\em Step 10}. The angle-compensated $\overset{\curvearrowright}{\mathbf{F}}_{k-1}(\theta, \rho)$ is then converted from polar to Cartesian coordinates, thus becoming $\overset{\curvearrowright}{\mathbf{F}}_{k-1}(x, y)$, in preparation for the comparison with $\mathbf{F}_{k}(x, y)$.
    \item {\em Step 11}. After Steps 9 and 10, $\mathbf{F}_{k}(x, y)$ is (ideally) a translated replica of $\overset{\curvearrowright}{\mathbf{F}}_{k-1}(x, y)$. The estimation $(\tilde{dx}, \tilde{dy})$ of the translation offset $(dx,dy)$ can now be derived following the same procedure adopted to estimate the rotation angle $d\theta$, starting from Step 5 to Step 8. To avoid confusing the reader with an overly complicated diagram, these steps have been incorporated into the \emph{Translational Offset Estimator} block of Fig.\ref{fig:Fourier_Mellin_Processing}, whose inputs,  however, are $\mathbf{F}_{k}(x, y)$ and $\overset{\curvearrowright}{\mathbf{F}}_{k-1}(x, y)$. 
    
\end{itemize}
%\begin{figure*}
%\begin{center}
	%\includegraphics[width=0.60\linewidth]{Figures/Approx_Fourier_Mellin}
	%\caption{Approximate Fourier-Mellin algorithm}
	%\label{fig:Approx_Fourier_Mellin_Processing}
%\end{center}
%\end{figure*}

\paragraph{Simplified Fourier-Mellin Algorithm} 
%{\color{red} Ci troverei un altro nome in quanto del Fourier-Melling non ha piu' niente essendo sparite le trasformate di Fourier ... non ho idee sul nome.   }
Having in mind that the \ac{THz} \ac{R-SLAM} is meant as an additional feature of future portable devices, the need to keep the computational effort, and consequently the energy consumption, as low as possible, suggested us to design a simpler, in principle less accurate, version of the above described Fourier-Mellin-based algorithm. The basic idea of this new version is to make the rotation-angle estimator (see the dashed red box in Fig.\ref{fig:Fourier_Mellin_Processing}) work directly on $\mathbf{F}_{k}(\theta, \rho)$ and $\mathbf{F}_{k-1}(\theta, \rho)$, rather than on the magnitude of their Fourier transforms. 
This means that the steps from 1 to 4 of the original algorithm are skipped, thus leading to the much simpler block scheme shown in Fig.\ref{fig:Approx_Fourier_Mellin_Processing}.
The assumption that makes the simplified algorithm sufficiently accurate is that, passing from $\mathbf{F}_{k-1}(\theta, \rho)$ to $\mathbf{F}_{k}(\theta, \rho)$, the impact of translations is much lower than the impact of rotations, so that the \emph{Rotation Angle Estimator} is not significantly affected by translations. 
This requires that a sufficiently small sampling time $T_{\text{s}}$ is chosen.
Despite its simplicity,  in the numerical results section it will be seen that the simplified algorithm has similar, sometimes better, performance than the  Fourier-Mellin-based algorithm.  
%This can be ascribed to the fact that the Fourier-Mellin-based algorithm is optimal only when $\mathbf{F}_k$ is a perfectly rotated and translated version of $\mathbf{F}_{k-1}$. In the presence of noise and other artifacts, typical of radar images, makes the algorithm more prune to errors because of the larger number of processing steps compared to the simplified algorithm. {\color{red} Decidere se questa frase giocarsela nei numerical results } 

\begin{figure*}[t]
\begin{center}
\centering
\subfigure[]{\includegraphics[width=0.38\linewidth]{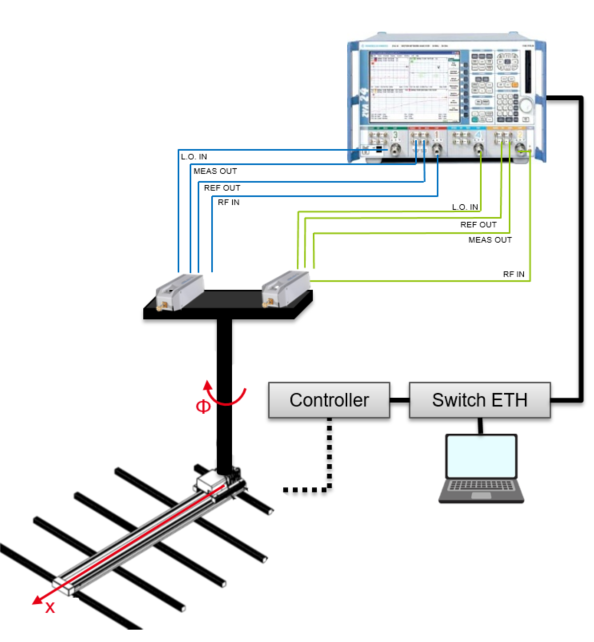}\label{fig:Setup}}
\subfigure[]{\includegraphics[width=0.52\linewidth]{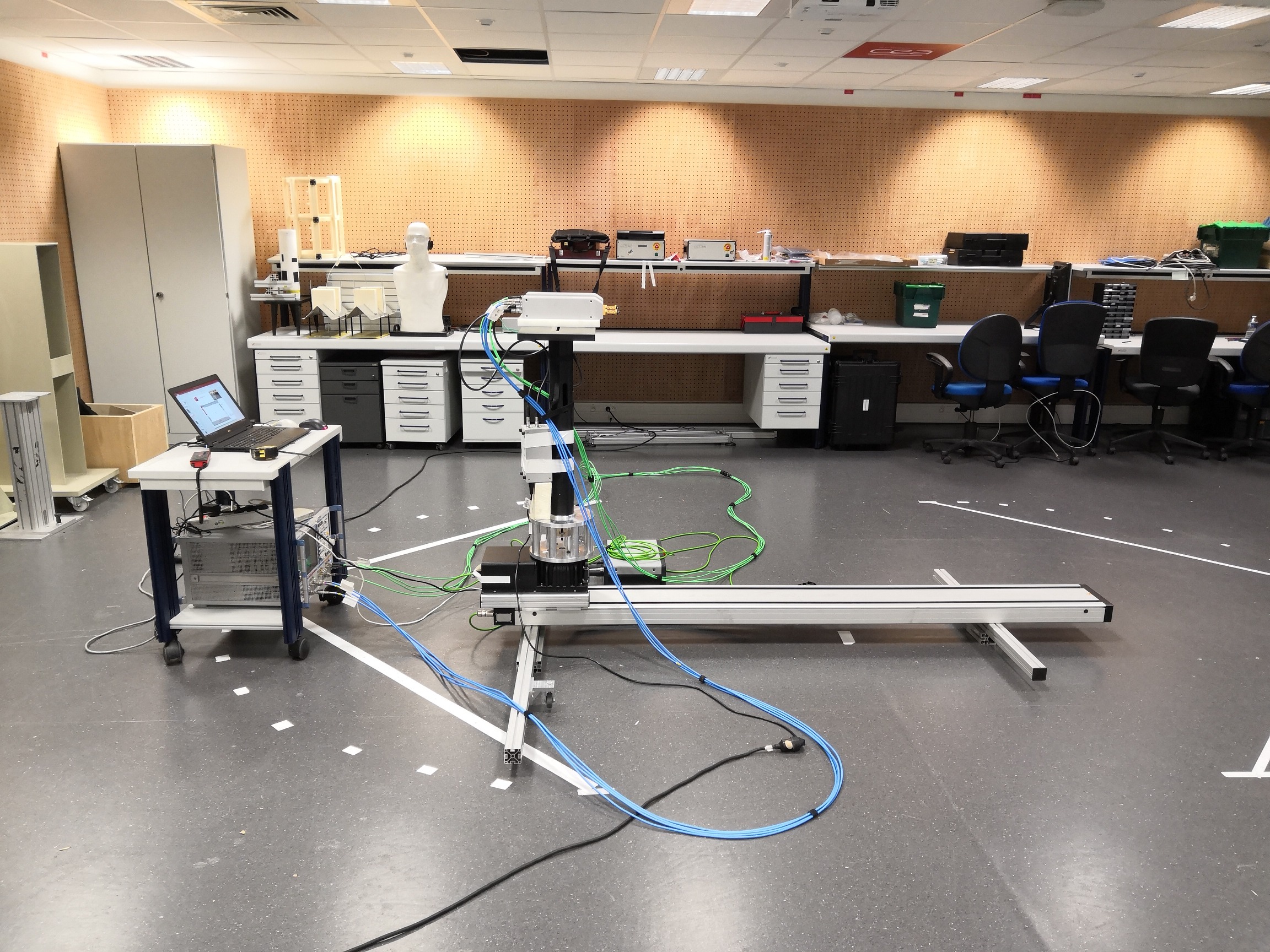}\label{fig:Layout}}
\caption{Measurement setup scheme (a) and picture of the environment (b).}
\end{center}
\end{figure*}

\section{Indoor Backscattering Characterization in the THz Band}
\label{Sec:Measures}

In this section, the measurement campaigns at \ac{THz} frequencies as well as the related processing are described. The real-world data collected during these  campaigns were subsequently used by the \ac{R-SLAM} algorithms, whose outcomes are presented in Sec.~\ref{sec:slamperformance}.
%in order to validate the \ac{THz} \ac{R-SLAM}.  

\subsection{Measurement Set Up}
To emulate the radar operation, the measurement setup, shown in Fig.~\ref{fig:Setup}, was based on a four-ports \ac{VNA} together with mm-Wave converters in order to cover the [$235-320$] $\mathrm{GHz}$ band. 
We employed two linearly polarized horn antennas with gain ${G_{\text{TX}} = G_{\text{RX}} = 20\, \mathrm{dBi}}$ and \ac{HPBW} of $18^\circ$. 
% The aperture size of the horn antennas is $4.30 \times 3 \ \text{mm}^2$.
The backscatter channel acquisition was operated in a quasi-monostatic configuration, as the TX and RX were co-located 
% The coupling can be quantified by filtering only the first part of the impulse response  and evaluating this window in the frequency: the value obtained is around -110 dB.
on a linear~-~angular positioner, which allowed mechanical steering and displacement over a 2-meter-long  X-axis. 
An external computer, connected via Ethernet cables to both the \ac{VNA} and the positioner controller, managed the acquisitions. 

%The frequency sweep step was set to $10\, \mathrm{MHz}$. 
% \red{(GP: Nella sezione IV.C si parla di 10 MHz)} \RD{credo che ci sia stata una confusione tra la IF bandwidth a 100 Hz, che é il filtro del VNA e determina il noise floor et lo step frequenziale à 10 MHz }%The power at the VNA's output ports is set between 10 to 12 dBm depending on the cable's attenuation at the measurement frequency.

\subsection{Indoor Measurement Campaigns}

\begin{figure}[t]
\centering
\subfigure[]{\includegraphics[width=1\linewidth]{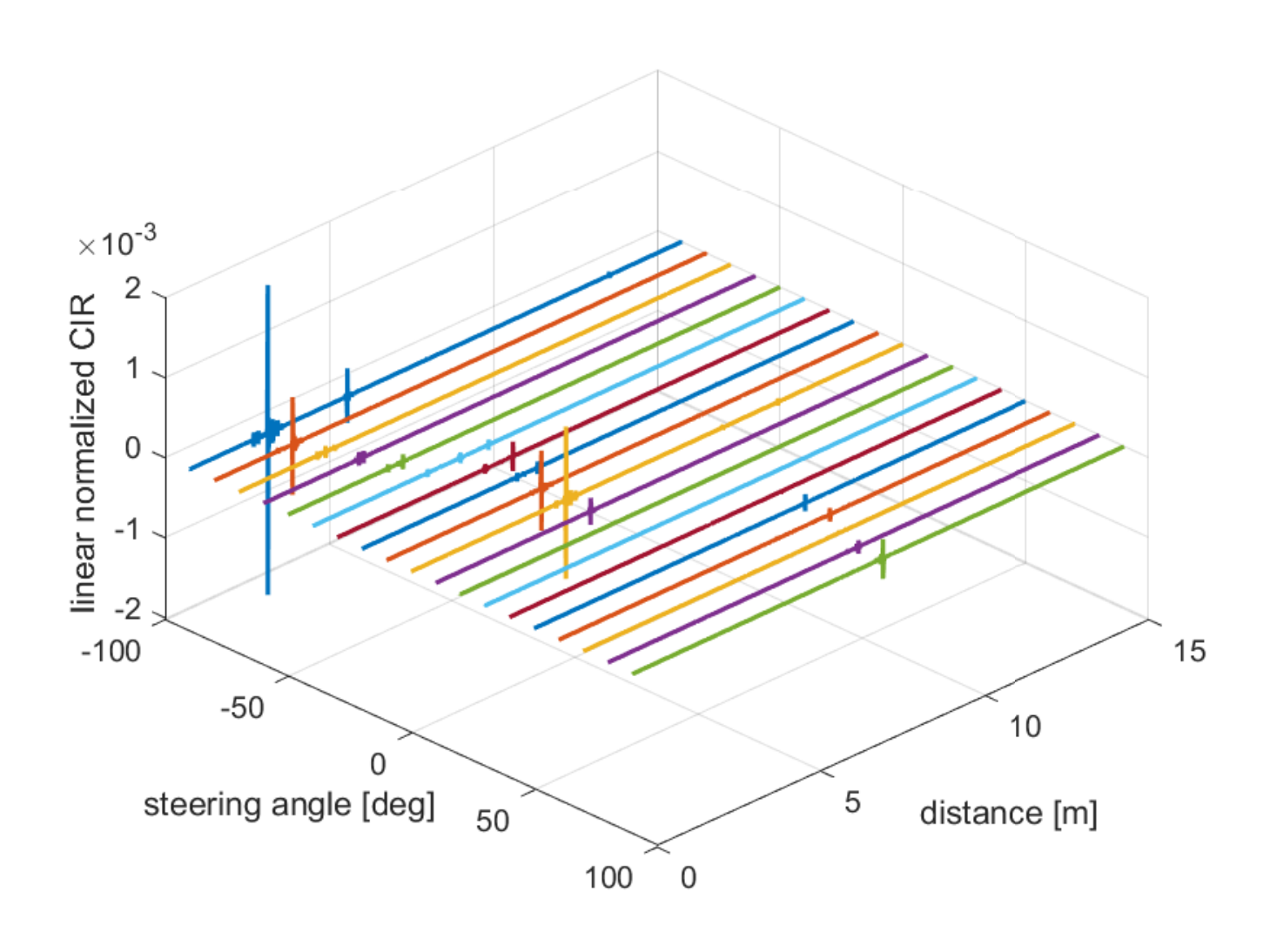}}
\subfigure[]{\includegraphics[width=1\linewidth]{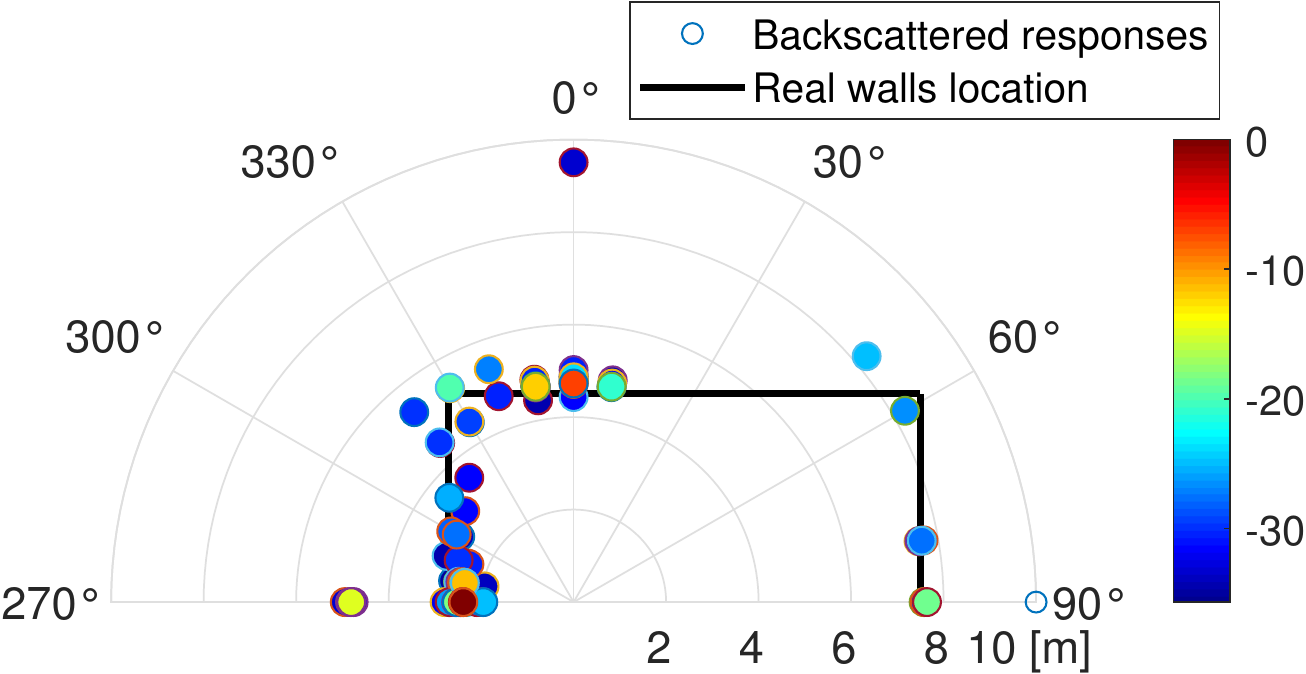}}
\caption{Normalized \acp{CIR} for all steering angles in position \#15 of Campaign C (a), with delays mapped into distances; results with the \ac{GEM} algorithm compared with the actual map of the environment (b). 
}\label{fig:Clean_pos15}
\end{figure} 

Two measurement campaigns were carried out at CEA-Leti in a laboratory/office room (shown in Fig.~\ref{fig:Layout}) with a size of ${10.2 \times 8.6\  \mathrm{m}^2}$. % at \ac{THz} frequencies.
%During the measurement campaign, the radar positioner performed a 2-D spatial grid of x-$\phi$ for each radar position. The measurements in the azimuth plane were taken each 10$^\circ$ according to the antennas HPBW, while in the X-axis the step size was dependent on the position of the radar. 
In the first measurement campaign, the radar scanned the laboratory in nine positions along a 2-meters straight path with a X-axis step of $0.25\, \mathrm{m}$. %, as shown in Fig.~\ref{fig:Rectangular_room_linear}.
This measurement was repeated twice, first with the radar pointed towards the direction of movement (Scenario A), then with the radar pointed perpendicular to the direction of movement (Scenario B). 

During the second measurement campaign (Scenario C), the radar scanned the environment in $46$ positions along an oval path, whose diameters were $5\, \mathrm{m}$ and $3\, \mathrm{m}$. The X-axis step size in the straight segments was of $0.40\,\mathrm{m}$, while in the curve the steps ranged from $0.23$ to $0.27\,\mathrm{m}$ so that the correlation between the channel responses of two adjacent positions was not lost.
In each location, the positioner rotated the antennas with step of $10^\circ$ in the steering interval $[-90^\circ, 90^\circ]$. To make it easy to read, the pictures of the three scenarios are reported together with the \ac{R-SLAM} performance in Sec.~\ref{sec:slamperformance}.

%The real-world data were collected in several positions, depending on the campaign, and for each steering angle, ranging in the interval [-90$^\circ$, 90$^\circ$] with  step of 10$^\circ$, 
\subsection{Generation of the Angle-Delay Matrix}
The above-described set up emulates a \acp{SFCW} radar. The measurement acquisition from $235$ to $320\, \mathrm{GHz}$, with step of $10\, \mathrm{MHz}$, for each position and steering direction corresponds to the \acp{CFR} including the antenna factor. 

%(one \ac{CFR} for each steering angle in each position).
%In the second, the radar covers a distance of 2 meters along a straight line and the step along the X-axis was fixed at 25 cm. 
%Fig.~\ref{fig:Rectangular_room_linear} and Fig.~\ref{fig:Rectangular_room_oval} show the radar positions for the straight and oval paths, respectively.
The \acp{CFR} acquired have been filtered in the frequency domain to reduce side lobe ringing and the real channel impulse responses (\acp{CIR}) were computed through an inverse FFT  with time resolution ${T_{\text{s}} = 1.56}\,\mathrm{ps}$.  
In any given position, the \acp{CIR} for 181 steering angles $\phi_n$ with step $1^\circ$ were computed by means of an interpolation process and subsequently collected to form the Angle-Delay matrix $\mathbf{H}$. 
 
%, and then converted into the Angle-Range matrix $\mathbf{AR}$ \red{ML:$\mathbf{F}$?}, as %described in Sec.\ref{Sec:Preprocessing} and shown in Fig. \ref{fig:Processing_chain}.
%This results into a  backscattered signal definition with delay starting from
% $T_{\text{min}}=T_{\text{s}}$, and delay resolution ${T_{\text{s}} = 1/2 f_{\text{max}} = 1.56}\,$ps. 

\subsection{Backscattering Channel Characterization}
Fig.~\ref{fig:Clean_pos15}(a) shows an example of normalized \acp{CIR} for different steering angles collected in position $15$ of Campaign C from which information about backscattering of the surrounding environment can be obtained. 
%of angular -dependent \acp{CIR}, with delays translated into distances, 
%to provide information about backscattering in the surrounding environment. %The CIRs' value was normalized by the number of points in frequency $M$ and the number of samples added in the zero-padding. 
% The peaks are the reflections of the sine tones transmitted due to the walls and obstacles present in front and side of the radar, as reported in Fig.~\ref{fig:Layout}. The peak in $\phi = 0^\circ $, in fact, corresponds to the distance of 4.72 m.\\
%As explained in Sec.\ref{Section:clean}, further CLEAN processing is necessary to mitigate the effect of the unwanted reflections. 
The \ac{GEM} algorithm (described in Sec.\ref{Section:clean}) was used and a map of the environment related to the current location was initially obtained by simply combining the power delay profiles (PDPs) associated with the \acp{CIR} in all the steering directions. 
% for one radar position at the time. 
The PDPs are compared with the location of the real walls in Fig.~\ref{fig:Clean_pos15}, visually compensating the curvature effect of the polar plot. Each point corresponds to a detected path, and the color represents the power normalized with respect to the highest path power. 
%
% Since the GEM detection threshold  was set to $-100\,\mathrm{dB}$, then also some weak double-reflections are present.  
%
It can be seen that the echoes detected by the radar are consistent with those expected from the main scatterers in the environment, i.e., the walls.

\begin{figure}[t]
\begin{center}
\centering
    % This file was created by matlab2tikz.
%
%The latest updates can be retrieved from
%  http://www.mathworks.com/matlabcentral/fileexchange/22022-matlab2tikz-matlab2tikz
%where you can also make suggestions and rate matlab2tikz.
%
\begin{tikzpicture}

\begin{axis}[%
%width=9.6in,
%height=7.885in,
%at={(0.969in,0.9in)},
%scale only axis,
width=0.35\linewidth, 
font=\footnotesize,
scale only axis,
point meta min=0,
point meta max=1,
xmin=5,
xmax=13,
xlabel style={font=\color{white!15!black}, font=\footnotesize},
xlabel={$\sigma_\tau$ [ns]},
ymin=0,
ymax=1,
ylabel style={font=\color{white!15!black}, font=\footnotesize, yshift=-1em},
ylabel={ECDF},
axis background/.style={fill=white},
xmajorgrids,
ymajorgrids,
legend style={at={(0.25,0.38)}, anchor=north west, legend cell align=left, align=left, draw=white!15!black, font=\footnotesize}
]
\addplot [color=blue, line width=1.5pt]
  table[row sep=crcr]{%
5.38969827898773	0\\
5.38969827898773	0.111111111111111\\
6.95939179985815	0.111111111111111\\
6.95939179985815	0.222222222222222\\
7.20383416048756	0.222222222222222\\
7.20383416048756	0.333333333333333\\
8.2621378033935	0.333333333333333\\
8.2621378033935	0.444444444444444\\
8.86737191031653	0.444444444444444\\
8.86737191031653	0.555555555555556\\
9.5041186847967	0.555555555555556\\
9.5041186847967	0.666666666666667\\
9.57466237087183	0.666666666666667\\
9.57466237087183	0.777777777777778\\
11.0622309344843	0.777777777777778\\
11.0622309344843	0.888888888888889\\
12.248176168109	0.888888888888889\\
12.248176168109	1\\
};
\addlegendentry{ECDF}

\addplot [color=black, dashed, line width=2.0pt]
  table[row sep=crcr]{%
5.32111350009652	0.0297675302075337\\
5.3917766056208	0.0334904164791962\\
5.46243971114508	0.0375325339844887\\
5.53310281666936	0.0419049510219654\\
5.60376592219364	0.0466176923749542\\
5.67442902771792	0.0516796492668586\\
5.7450921332422	0.0570984998050415\\
5.81575523876648	0.0628806406264944\\
5.88641834429076	0.0690311302729436\\
5.95708144981504	0.0755536446396021\\
6.02774455533932	0.0824504446637625\\
6.0984076608636	0.0897223562497237\\
6.16907076638788	0.097368762267617\\
6.23973387191216	0.105387606317503\\
6.31039697743643	0.113775407818204\\
6.38106008296071	0.122527287863779\\
6.45172318848499	0.131637005190087\\
6.52238629400927	0.141097001509722\\
6.59304939953355	0.150898455405848\\
6.66371250505783	0.161031343923636\\
6.73437561058211	0.171484510961627\\
6.80503871610639	0.182245741543554\\
6.87570182163067	0.193301841042988\\
6.94636492715495	0.204638718437522\\
7.01702803267923	0.216241472684856\\
7.08769113820351	0.228094481338807\\
7.15835424372779	0.240181490557723\\
7.22901734925207	0.252485705699587\\
7.29968045477635	0.26498988174613\\
7.37034356030063	0.277676412851155\\
7.44100666582491	0.290527420364978\\
7.51166977134918	0.303524838746114\\
7.58233287687346	0.316650498832261\\
7.65299598239774	0.329886208004172\\
7.72365908792202	0.343213826837376\\
7.7943221934463	0.356615341897176\\
7.86498529897058	0.370072934391205\\
7.93564840449486	0.383569044450613\\
8.00631151001914	0.397086430865061\\
8.07697461554342	0.410608226147938\\
8.1476377210677	0.424117986856098\\
8.21830082659198	0.437599739132845\\
8.28896393211626	0.451038019483723\\
8.35962703764054	0.464417910831704\\
8.43029014316482	0.477725073931718\\
8.5009532486891	0.490945774254037\\
8.57161635421338	0.504066904471987\\
8.64227945973766	0.5170760027118\\
8.71294256526194	0.529961266741347\\
8.78360567078622	0.542711564290191\\
8.85426877631049	0.555316439705918\\
8.92493188183477	0.567766117161389\\
8.99559498735905	0.58005150063452\\
9.06625809288333	0.592164170886683\\
9.13692119840761	0.604096379667996\\
9.20758430393189	0.615841041377963\\
9.27824740945617	0.627391722408193\\
9.34891051498045	0.638742628390607\\
9.41957362050473	0.649888589569789\\
9.49023672602901	0.660825044512078\\
9.56089983155329	0.671548022356988\\
9.63156293707757	0.682054123808527\\
9.70222604260185	0.692340501055339\\
9.77288914812613	0.702404836799307\\
9.84355225365041	0.712245322562611\\
9.91421535917469	0.721860636433165\\
9.98487846469897	0.731249920398236\\
10.0555415702232	0.74041275740573\\
10.1262046757475	0.749349148282358\\
10.1968677812718	0.758059488627737\\
10.2675308867961	0.766544545793421\\
10.3381939923204	0.774805436046078\\
10.4088570978446	0.782843602004491\\
10.4795202033689	0.790660790430842\\
10.5501833088932	0.798259030447908\\
10.6208464144175	0.805640612245304\\
10.6915095199418	0.812808066329863\\
10.762172625466	0.8197641433676\\
10.8328357309903	0.826511794657514\\
10.9034988365146	0.83305415327072\\
10.9741619420389	0.839394515882079\\
11.0448250475632	0.845536325315637\\
11.1154881530874	0.85148315381975\\
11.1861512586117	0.857238687082732\\
11.256814364136	0.862806708995314\\
11.3274774696603	0.868191087162\\
11.3981405751846	0.873395759159594\\
11.4688036807088	0.878424719537779\\
11.5394667862331	0.883282007553558\\
11.6101298917574	0.887971695628634\\
11.6807929972817	0.89249787851642\\
11.751456102806	0.896864663163278\\
11.8221192083302	0.901076159246755\\
11.8927823138545	0.905136470372066\\
11.9634454193788	0.909049685906766\\
12.0341085249031	0.912819873432483\\
12.1047716304274	0.916451071791757\\
12.1754347359516	0.919947284707334\\
12.2460978414759	0.923312474950843\\
12.3167609470002	0.926550559037399\\
};
\addlegendentry{Log-normal}

\end{axis}
\end{tikzpicture}%
    % This file was created by matlab2tikz.
%
%The latest updates can be retrieved from
%  http://www.mathworks.com/matlabcentral/fileexchange/22022-matlab2tikz-matlab2tikz
%where you can also make suggestions and rate matlab2tikz.
%
\begin{tikzpicture}

\begin{axis}[%
width=0.35\linewidth, 
font=\footnotesize,
scale only axis,
point meta min=0,
point meta max=1,
xmin=40,
xmax=90,
xtick={30,40,50,60,70,80,90},
xlabel style={font=\color{white!15!black}, font=\footnotesize},
xlabel={$\sigma_\phi$ [$^\circ$]},
ymin=0,
ymax=1,
ylabel style={font=\color{white!15!black}, font=\footnotesize, yshift=-1em},
ylabel={ECDF},
axis background/.style={fill=white},
xmajorgrids,
ymajorgrids,
font=\small,
legend style={at={(0.25,0.38)}, anchor=north west, legend cell align=left, align=left, draw=white!15!black, font=\footnotesize}
]
\addplot [color=red, line width=1.5pt]
  table[row sep=crcr]{%
44.6908907985469	0\\
44.6908907985469	0.111111111111111\\
52.2174425785785	0.111111111111111\\
52.2174425785785	0.222222222222222\\
58.2239790754046	0.222222222222222\\
58.2239790754046	0.333333333333333\\
63.4180495236413	0.333333333333333\\
63.4180495236413	0.444444444444444\\
63.4738363662294	0.444444444444444\\
63.4738363662294	0.555555555555556\\
63.7295956897096	0.555555555555556\\
63.7295956897096	0.666666666666667\\
64.918975750366	0.666666666666667\\
64.918975750366	0.777777777777778\\
65.1556396938955	0.777777777777778\\
65.1556396938955	0.888888888888889\\
77.97782429255	0.888888888888889\\
77.97782429255	1\\
};
\addlegendentry{ECDF}

\addplot [color=black, dashed, line width=2.0pt]
  table[row sep=crcr]{%
44.3580214636068	0.0216865496614739\\
44.7009777480905	0.0243636003741912\\
45.0439340325742	0.0272891566188353\\
45.3868903170578	0.0304765736326118\\
45.7298466015415	0.0339389605319261\\
46.0728028860252	0.0376890837357825\\
46.4157591705088	0.0417392695593365\\
46.7587154549925	0.0461013069905567\\
47.1016717394762	0.0507863516740992\\
47.4446280239598	0.0558048321225048\\
47.7875843084435	0.0611663591558792\\
48.1305405929272	0.0668796395378678\\
48.4734968774108	0.0729523947287392\\
48.8164531618945	0.0793912856167641\\
49.1594094463782	0.0862018440180019\\
49.5023657308619	0.0933884116535184\\
49.8453220153455	0.100954087223378\\
50.1882782998292	0.108900682100213\\
50.5312345843129	0.117228685063293\\
50.8741908687965	0.125937236388695\\
51.2171471532802	0.135024111503867\\
51.5601034377639	0.144485714307495\\
51.9030597222475	0.154317080149475\\
52.2460160067312	0.164511888362651\\
52.5889722912149	0.175062484139043\\
52.9319285756985	0.18595990944987\\
53.2748848601822	0.197193942621862\\
53.6178411446659	0.208753146103073\\
53.9607974291495	0.22062492188041\\
54.3037537136332	0.232795573949018\\
54.6467099981169	0.245250377180914\\
54.9896662826005	0.257973651897098\\
55.3326225670842	0.27094884341391\\
55.6755788515679	0.284158605810623\\
56.0185351360516	0.297584889150885\\
56.3614914205352	0.311209029385523\\
56.7044477050189	0.325011840167666\\
57.0474039895026	0.338973705823051\\
57.3903602739862	0.353074674737628\\
57.7333165584699	0.367294552450959\\
58.0762728429536	0.381612993776378\\
58.4192291274372	0.396009593306859\\
58.7621854119209	0.410463973708161\\
59.1051416964046	0.424955871247292\\
59.4480979808882	0.439465218053924\\
59.7910542653719	0.453972220664302\\
60.1340105498556	0.468457434450594\\
60.4769668343392	0.482901833593003\\
60.8199231188229	0.497286876306355\\
61.1628794033066	0.511594565087039\\
61.5058356877902	0.525807501799105\\
61.8487919722739	0.539908937469874\\
62.1917482567576	0.553882816714957\\
62.5347045412413	0.567713816759697\\
62.8776608257249	0.581387381068457\\
63.2206171102086	0.59488974763473\\
63.5635733946923	0.608207972023118\\
63.9065296791759	0.621329945289276\\
64.2494859636596	0.634244406935172\\
64.5924422481433	0.646940953084953\\
64.9353985326269	0.659410040090959\\
65.2783548171106	0.671642983800262\\
65.6213111015943	0.683631954729375\\
65.9642673860779	0.695369969408786\\
66.3072236705616	0.706850878169669\\
66.6501799550453	0.718069349652825\\
66.9931362395289	0.729020852324602\\
67.3360925240126	0.739701633286632\\
67.6790488084963	0.750108694665697\\
68.02200509298	0.760239767867296\\
68.3649613774636	0.770093285971581\\
68.7079176619473	0.779668354543633\\
69.050873946431	0.788964721121625\\
69.3938302309146	0.797982743636628\\
69.7367865153983	0.806723358006793\\
70.079742799882	0.815188045136541\\
70.4226990843656	0.823378797538544\\
70.7656553688493	0.831298085782777\\
71.108611653333	0.838948824962885\\
71.4515679378166	0.846334341355892\\
71.7945242223003	0.853458339436803\\
72.137480506784	0.860324869395192\\
72.4804367912676	0.866938295286578\\
72.8233930757513	0.87330326393723\\
73.166349360235	0.879424674707246\\
73.5093056447186	0.885307650203392\\
73.8522619292023	0.890957508020228\\
74.195218213686	0.89637973357569\\
74.5381744981696	0.901579954095539\\
74.8811307826533	0.906563913789896\\
75.224087067137	0.911337450254626\\
75.5670433516207	0.915906472120502\\
75.9099996361043	0.920276937964029\\
76.252955920588	0.924454836485317\\
76.5959122050717	0.928446167950802\\
76.9388684895553	0.932256926891516\\
77.281824774039	0.935893086041358\\
77.6247810585227	0.939360581494111\\
77.9677373430063	0.942665299052997\\
78.31069362749	0.945813061742131\\
};
\addlegendentry{Log-normal}

\end{axis}

\end{tikzpicture}%
    % This file was created by matlab2tikz.
%
%The latest updates can be retrieved from
%  http://www.mathworks.com/matlabcentral/fileexchange/22022-matlab2tikz-matlab2tikz
%where you can also make suggestions and rate matlab2tikz.
%
\begin{tikzpicture}

\begin{axis}[%
%width=9.111in,
%height=7.496in,
%at={(1.424in,1.17in)},
%scale only axis,
width=0.35\linewidth, 
scale only axis,
point meta min=0,
point meta max=1,
font=\footnotesize,
xmin=0,
xmax=14,
xlabel style={font=\color{white!15!black},font=\footnotesize},
xlabel={$\sigma_\tau$ [ns]},
ymin=0,
ymax=1,
ylabel style={font=\color{white!15!black}, font=\footnotesize, yshift=-1em},
ylabel={ECDF},
axis background/.style={fill=white},
xmajorgrids,
ymajorgrids,
legend style={at={(0.25,0.38)}, anchor=north west, legend cell align=left, align=left, draw=white!15!black, font=\footnotesize}
]
\addplot [color=blue, line width=1.5pt]
  table[row sep=crcr]{%
2.06788591009724	0\\
2.06788591009724	0.111111111111111\\
2.28612053533674	0.111111111111111\\
2.28612053533674	0.222222222222222\\
3.24682548698668	0.222222222222222\\
3.24682548698668	0.333333333333333\\
4.20035309869088	0.333333333333333\\
4.20035309869088	0.444444444444444\\
5.29756347842544	0.444444444444444\\
5.29756347842544	0.555555555555556\\
5.82429531759204	0.555555555555556\\
5.82429531759204	0.666666666666667\\
7.1583425259123	0.666666666666667\\
7.1583425259123	0.777777777777778\\
8.47767637780195	0.777777777777778\\
8.47767637780195	0.888888888888889\\
12.4859663813698	0.888888888888889\\
12.4859663813698	1\\
};
\addlegendentry{ECDF}

\addplot [color=black, dashed, line width=2.0pt]
  table[row sep=crcr]{%
1.96370510538451	0.065326791893954\\
2.07104290417944	0.0774082570485604\\
2.17838070297437	0.0903823053929058\\
2.2857185017693	0.104170098334569\\
2.39305630056423	0.118689248563628\\
2.50039409935916	0.133855939371344\\
2.60773189815409	0.149586649933928\\
2.71506969694902	0.16579952529436\\
2.82240749574394	0.18241543425274\\
2.92974529453887	0.199358758850556\\
3.0370830933338	0.21655795714542\\
3.14442089212873	0.233945937589494\\
3.25175869092366	0.251460279269211\\
3.35909648971859	0.269043328007833\\
3.46643428851352	0.286642194171539\\
3.57377208730845	0.304208674127417\\
3.68110988610338	0.321699113769063\\
3.78844768489831	0.339074229389235\\
3.89578548369323	0.35629889844152\\
4.00312328248816	0.373341930375783\\
4.11046108128309	0.390175825725259\\
4.21779888007802	0.406776529930922\\
4.32513667887295	0.423123186974376\\
4.43247447766788	0.439197896717624\\
4.53981227646281	0.454985478883061\\
4.64715007525774	0.470473245819092\\
4.75448787405267	0.485650785558488\\
4.8618256728476	0.500509756163949\\
4.96916347164253	0.515043691947481\\
5.07650127043745	0.529247821829177\\
5.18383906923238	0.543118899851606\\
5.29117686802731	0.556655047675306\\
5.39851466682224	0.569855608738021\\
5.50585246561717	0.582721013656177\\
5.6131902644121	0.595252656374157\\
5.72052806320703	0.60745278051886\\
5.82786586200196	0.619324375388664\\
5.93520366079689	0.630871080993037\\
6.04254145959182	0.642097101558058\\
6.14987925838675	0.653007126921313\\
6.25721705718167	0.6636062612546\\
6.3645548559766	0.673899958572853\\
6.47189265477153	0.683893964511079\\
6.57923045356646	0.693594263876839\\
6.68656825236139	0.703007033512844\\
6.79390605115632	0.712138600031888\\
6.90124384995125	0.720995402014083\\
7.00858164874618	0.729583956283683\\
7.11591944754111	0.737910827909418\\
7.22325724633603	0.745982603597968\\
7.33059504513096	0.753805868174779\\
7.43793284392589	0.761387183869857\\
7.54527064272082	0.768733072148249\\
7.65260844151575	0.775849997845778\\
7.75994624031068	0.782744355390086\\
7.86728403910561	0.789422456905315\\
7.97462183790054	0.795890522015672\\
8.08195963669547	0.802154669178974\\
8.1892974354904	0.808220908395823\\
8.29663523428532	0.814095135153601\\
8.40397303308025	0.819783125476958\\
8.51131083187518	0.825290531967942\\
8.61864863067011	0.830622880729505\\
8.72598642946504	0.835785569075817\\
8.83332422825997	0.840783863941773\\
8.9406620270549	0.84562290091219\\
9.04799982584983	0.850307683798717\\
9.15533762464476	0.854843084699305\\
9.26267542343969	0.859233844481317\\
9.37001322223461	0.863484573635096\\
9.47735102102954	0.867599753449982\\
9.58468881982447	0.871583737469536\\
9.6920266186194	0.875440753187026\\
9.79936441741433	0.879174903946195\\
9.90670221620926	0.882790171015856\\
10.0140400150042	0.886290415810172\\
10.1213778137991	0.889679382229362\\
10.228715612594	0.892960699098316\\
10.336053411389	0.896137882682981\\
10.4433912101839	0.899214339266602\\
10.5507290089788	0.902193367769912\\
10.6580668077738	0.905078162401123\\
10.7654046065687	0.907871815323261\\
10.8727424053636	0.910577319327804\\
10.9800802041586	0.913197570504956\\
11.0874180029535	0.915735370902074\\
11.1947558017484	0.91819343116284\\
11.3020936005433	0.920574373140759\\
11.4094313993383	0.922880732481441\\
11.5167691981332	0.925114961168888\\
11.6241069969281	0.927279430031753\\
11.7314447957231	0.929376431206141\\
11.838782594518	0.931408180552098\\
11.9461203933129	0.933376820021456\\
12.0534581921078	0.935284419975128\\
12.1607959909028	0.937132981448372\\
12.2681337896977	0.938924438362901\\
12.3754715884926	0.940660659685021\\
12.4828093872876	0.942343451529264\\
12.5901471860825	0.943974559207249\\
};
\addlegendentry{Log-normal}

\end{axis}
\end{tikzpicture}%
    % This file was created by matlab2tikz.
%
%The latest updates can be retrieved from
%  http://www.mathworks.com/matlabcentral/fileexchange/22022-matlab2tikz-matlab2tikz
%where you can also make suggestions and rate matlab2tikz.
%
\begin{tikzpicture}

\begin{axis}[%
%width=9.229in,
%height=7.532in,
%at={(1.306in,1.17in)},
%scale only axis,
width=0.35\linewidth, 
scale only axis,
font=\footnotesize,
point meta min=0,
point meta max=1,
xmin=65,
xmax=90,
xlabel style={font=\color{white!15!black}, font=\small},
xlabel={$\sigma_\phi$ [$^\circ$]},
ymin=0,
ymax=1,
ylabel style={font=\color{white!15!black}, font=\footnotesize, yshift=-1em},
ylabel={ECDF},
axis background/.style={fill=white},
xmajorgrids,
ymajorgrids,
legend style={at={(0.25,0.38)}, anchor=north west, legend cell align=left, align=left, draw=white!15!black, font=\footnotesize}
]
\addplot [color=red, line width=1.5pt]
  table[row sep=crcr]{%
68.5887456401419	0\\
68.5887456401419	0.111111111111111\\
70.1209751559355	0.111111111111111\\
70.1209751559355	0.222222222222222\\
71.2627334196837	0.222222222222222\\
71.2627334196837	0.333333333333333\\
74.3669250390077	0.333333333333333\\
74.3669250390077	0.444444444444444\\
74.7414026918362	0.444444444444444\\
74.7414026918362	0.555555555555556\\
77.6876252224465	0.555555555555556\\
77.6876252224465	0.666666666666667\\
80.0000208871436	0.666666666666667\\
80.0000208871436	0.777777777777778\\
80.6303657151071	0.777777777777778\\
80.6303657151071	0.888888888888889\\
90.4729775965278	0.888888888888889\\
90.4729775965278	1\\
};
\addlegendentry{ECFD}

\addplot [color=black, dashed, line width=2.0pt]
  table[row sep=crcr]{%
68.3699033205781	0.103743980764433\\
68.5953772255833	0.110831790316813\\
68.8208511305885	0.118235050246819\\
69.0463250355936	0.125955407438481\\
69.2717989405988	0.133993663645985\\
69.497272845604	0.142349749397429\\
69.7227467506092	0.151022702264616\\
69.9482206556144	0.160010649718838\\
70.1736945606196	0.169310796756047\\
70.3991684656248	0.178919418436407\\
70.62464237063	0.188831857443228\\
70.8501162756351	0.199042526725273\\
71.0755901806403	0.209544917244755\\
71.3010640856455	0.220331610811638\\
71.5265379906507	0.231394297943295\\
71.7520118956559	0.24272380064796\\
71.9774858006611	0.254310099990806\\
72.2029597056663	0.266142368263658\\
72.4284336106715	0.278209005543404\\
72.6539075156767	0.290497680390535\\
72.8793814206818	0.302995374408473\\
73.104855325687	0.315688430356245\\
73.3303292306922	0.328562603482418\\
73.5558031356974	0.341603115726735\\
73.7812770407026	0.3547947124181\\
74.0067509457078	0.368121721083361\\
74.232224850713	0.381568111970888\\
74.4576987557181	0.395117559886091\\
74.6831726607233	0.408753506933158\\
74.9086465657285	0.422459225757709\\
75.1341204707337	0.43621788288935\\
75.3595943757389	0.450012601790504\\
75.5850682807441	0.463826525228813\\
75.8105421857493	0.477642876604015\\
76.0360160907545	0.491445019876798\\
76.2614899957596	0.505216517766295\\
76.4869639007648	0.518941187904104\\
76.71243780577	0.532603156656104\\
76.9379117107752	0.546186910348281\\
77.1633856157804	0.5596773436592\\
77.3888595207856	0.573059804969107\\
77.6143334257908	0.586320138483974\\
77.839807330796	0.599444722981292\\
78.0652812358012	0.612420507053513\\
78.2907551408063	0.625235040753509\\
78.5162290458115	0.637876503575139\\
78.7417029508167	0.6503337287295\\
78.9671768558219	0.662596223704614\\
79.1926507608271	0.674654187122097\\
79.4181246658323	0.686498521929128\\
79.6435985708375	0.698120844987129\\
79.8690724758427	0.709513493140616\\
80.0945463808478	0.720669525869395\\
80.320020285853	0.731582724645929\\
80.5454941908582	0.742247589135974\\
80.7709680958634	0.752659330395324\\
80.9964420008686	0.762813861228139\\
81.2219159058738	0.772707783883225\\
81.447389810879	0.782338375273394\\
81.6728637158842	0.791703569910178\\
81.8983376208894	0.800801940751294\\
82.1238115258945	0.809632678161679\\
82.3492854308997	0.818195567190658\\
82.5747593359049	0.826490963367934\\
82.8002332409101	0.834519767219661\\
83.0257071459153	0.842283397703135\\
83.2511810509205	0.849783764754424\\
83.4766549559257	0.857023241138108\\
83.7021288609309	0.864004633781851\\
83.927602765936	0.870731154771325\\
84.1530766709412	0.87720639217286\\
84.3785505759464	0.883434280842369\\
84.6040244809516	0.889419073369817\\
84.8294983859568	0.895165311298513\\
85.054972290962	0.900677796748434\\
85.2804461959672	0.905961564562256\\
85.5059201009724	0.911021855082241\\
85.7313940059775	0.915864087655436\\
85.9568679109827	0.92049383495408\\
86.1823418159879	0.924916798187607\\
86.4078157209931	0.929138783272398\\
86.6332896259983	0.933165678015315\\
86.8587635310035	0.937003430357478\\
87.0842374360087	0.940658027715314\\
87.3097113410139	0.944135477447038\\
87.535185246019	0.947441788464292\\
87.7606591510242	0.950582954000652\\
87.9861330560294	0.953564935541293\\
88.2116069610346	0.956393647911153\\
88.4370808660398	0.959074945512552\\
88.662554771045	0.961614609697367\\
88.8880286760502	0.964018337253554\\
89.1135025810554	0.966291729981068\\
89.3389764860605	0.968440285327989\\
89.5644503910657	0.970469388053945\\
89.7899242960709	0.972384302884746\\
90.0153982010761	0.974190168119403\\
90.2408721060813	0.975891990148497\\
90.4663460110865	0.977494638841006\\
90.6918199160917	0.979002843755376\\
};
\addlegendentry{Log-normal}

\end{axis}
\end{tikzpicture}%
    % This file was created by matlab2tikz.
%
%The latest updates can be retrieved from
%  http://www.mathworks.com/matlabcentral/fileexchange/22022-matlab2tikz-matlab2tikz
%where you can also make suggestions and rate matlab2tikz.
%
\begin{tikzpicture}

\begin{axis}[%
%width=9.011in,
%height=7.568in,
%at={(1.524in,1.17in)},
%scale only axis,
width=0.35\linewidth, 
scale only axis,
point meta min=0,
point meta max=1,
font=\footnotesize,
xmin=4,
xmax=20,
xtick={4,6,8,10,12,14,16,18,20},
xtick={5,10,15,20},
xlabel style={font=\color{white!15!black}, font=\footnotesize},
xlabel={$\sigma_\tau$ [ns]},
ymin=0,
ymax=1,
ylabel style={font=\color{white!15!black}, font=\footnotesize, yshift=-1em},
ylabel={ECDF},
axis background/.style={fill=white},
xmajorgrids,
ymajorgrids,
legend style={at={(0.25,0.38)}, anchor=north west, legend cell align=left, align=left, draw=white!15!black, font=\footnotesize}
]
\addplot [color=blue, line width=1.5pt]
  table[row sep=crcr]{%
4.21696070308551	0\\
4.21696070308551	0.0217391304347826\\
4.84395480535266	0.0217391304347826\\
4.84395480535266	0.0434782608695652\\
5.02581145570323	0.0434782608695652\\
5.02581145570323	0.0652173913043478\\
5.66458618531012	0.0652173913043478\\
5.66458618531012	0.0869565217391304\\
6.15118954387415	0.0869565217391304\\
6.15118954387415	0.108695652173913\\
6.18941048303224	0.108695652173913\\
6.18941048303224	0.130434782608696\\
6.23357730652098	0.130434782608696\\
6.23357730652098	0.152173913043478\\
7.06088457757027	0.152173913043478\\
7.06088457757027	0.173913043478261\\
7.27933152509235	0.173913043478261\\
7.27933152509235	0.195652173913043\\
7.64511468423523	0.195652173913043\\
7.64511468423523	0.217391304347826\\
8.09996592163318	0.217391304347826\\
8.09996592163318	0.239130434782609\\
8.58998200122788	0.239130434782609\\
8.58998200122788	0.260869565217391\\
8.86651201793753	0.260869565217391\\
8.86651201793753	0.282608695652174\\
8.96932566621524	0.282608695652174\\
8.96932566621524	0.304347826086956\\
9.6801977434984	0.304347826086956\\
9.6801977434984	0.326086956521739\\
9.92836737327201	0.326086956521739\\
9.92836737327201	0.347826086956522\\
10.1057531930417	0.347826086956522\\
10.1057531930417	0.369565217391304\\
10.5269869110792	0.369565217391304\\
10.5269869110792	0.391304347826087\\
10.5308878605366	0.391304347826087\\
10.5308878605366	0.413043478260869\\
11.4749549955122	0.413043478260869\\
11.4749549955122	0.434782608695652\\
11.8292930801238	0.434782608695652\\
11.8292930801238	0.456521739130434\\
11.8907975478688	0.456521739130434\\
11.8907975478688	0.478260869565217\\
12.1193829326009	0.478260869565217\\
12.1193829326009	0.5\\
12.2074693439478	0.5\\
12.2074693439478	0.521739130434782\\
12.5820040501377	0.521739130434782\\
12.5820040501377	0.543478260869565\\
12.7449038962288	0.543478260869565\\
12.7449038962288	0.565217391304347\\
12.7721238389995	0.565217391304347\\
12.7721238389995	0.58695652173913\\
12.8883750145817	0.58695652173913\\
12.8883750145817	0.608695652173913\\
12.9363824152747	0.608695652173913\\
12.9363824152747	0.630434782608695\\
13.0694860370428	0.630434782608695\\
13.0694860370428	0.652173913043478\\
13.2453150325941	0.652173913043478\\
13.2453150325941	0.673913043478261\\
13.2534562229276	0.673913043478261\\
13.2534562229276	0.695652173913043\\
13.7405025615157	0.695652173913043\\
13.7405025615157	0.717391304347826\\
14.0995630198093	0.717391304347826\\
14.0995630198093	0.739130434782608\\
14.3105706644466	0.739130434782608\\
14.3105706644466	0.760869565217391\\
14.7725114711556	0.760869565217391\\
14.7725114711556	0.782608695652174\\
15.0207515093046	0.782608695652174\\
15.0207515093046	0.804347826086956\\
15.3513114311512	0.804347826086956\\
15.3513114311512	0.826086956521739\\
15.5806543939221	0.826086956521739\\
15.5806543939221	0.847826086956522\\
16.0015560240785	0.847826086956522\\
16.0015560240785	0.869565217391304\\
16.9966007888218	0.869565217391304\\
16.9966007888218	0.891304347826087\\
17.2234798669383	0.891304347826087\\
17.2234798669383	0.91304347826087\\
17.6143513771452	0.91304347826087\\
17.6143513771452	0.934782608695652\\
18.2386639795118	0.934782608695652\\
18.2386639795118	0.956521739130435\\
19.8145693576275	0.956521739130435\\
19.8145693576275	0.978260869565217\\
20.4091101207395	0.978260869565217\\
20.4091101207395	1\\
};
\addlegendentry{ECDF}

\addplot [color=black, dashed, line width=2.0pt]
  table[row sep=crcr]{%
4.05503920890897	0.0060390309597043\\
4.22186741503025	0.00802931469312346\\
4.38869562115153	0.0104643181116999\\
4.55552382727282	0.0133928466857689\\
4.7223520333941	0.0168606565158007\\
4.88918023951539	0.0209094393230235\\
5.05600844563667	0.0255759660330732\\
5.22283665175795	0.0308914048042447\\
5.38966485787924	0.0368808194526206\\
5.55649306400052	0.0435628458820361\\
5.72332127012181	0.050949537476625\\
5.89014947624309	0.0590463654451813\\
6.05697768236437	0.0678523567027951\\
6.22380588848566	0.0773603498544524\\
6.39063409460694	0.0875573489928716\\
6.55746230072823	0.0984249551149334\\
6.72429050684951	0.109939855781375\\
6.89111871297079	0.122074354995046\\
7.05794691909208	0.134796926980465\\
7.22477512521336	0.148072779465365\\
7.39160333133465	0.161864414074142\\
7.55843153745593	0.176132173450257\\
7.72525974357721	0.190834766659368\\
7.8920879496985	0.205929766237142\\
8.05891615581978	0.221374071902259\\
8.22574436194106	0.237124337436627\\
8.39257256806235	0.253137358533079\\
8.55940077418363	0.26937042052583\\
8.72622898030492	0.285781605857139\\
8.8930571864262	0.302330061905418\\
9.05988539254749	0.318976230419222\\
9.22671359866877	0.335682040282941\\
9.39354180479005	0.352411065699679\\
9.56037001091133	0.369128652130243\\
9.72719821703262	0.385802012489858\\
9.8940264231539	0.402400296190233\\
10.0608546292752	0.418894633637402\\
10.2276828353965	0.435258158766997\\
10.3945110415178	0.451466012129238\\
10.561339247639	0.467495326935009\\
10.7281674537603	0.483325200350202\\
10.8949956598816	0.498936652184901\\
11.0618238660029	0.514312572972751\\
11.2286520721242	0.529437663278881\\
11.3954802782455	0.544298365915954\\
11.5623084843667	0.558882792590522\\
11.729136690488	0.573180646348237\\
11.8959648966093	0.587183141038736\\
12.0627931027306	0.600882918880282\\
12.2296213088519	0.614273967071861\\
12.3964495149732	0.62735153427663\\
12.5632777210944	0.64011204768614\\
12.7301059272157	0.652553031269379\\
12.896934133337	0.664673025714593\\
13.0637623394583	0.676471510484675\\
13.2305905455796	0.687948828328367\\
13.3974187517009	0.699106112519292\\
13.5642469578222	0.709945217032283\\
13.7310751639434	0.720468649811273\\
13.8979033700647	0.73067950923457\\
14.064731576186	0.740581423841104\\
14.2315597823073	0.750178495344745\\
14.3983879884286	0.759475244932531\\
14.5652161945499	0.768476562816078\\
14.7320444006711	0.777187660983177\\
14.8988726067924	0.785614029078092\\
15.0657008129137	0.793761393324042\\
15.232529019035	0.801635678389306\\
15.3993572251563	0.809242972088987\\
15.5661854312776	0.81658949280751\\
15.7330136373988	0.82368155952187\\
15.8998418435201	0.830525564302466\\
16.0666700496414	0.837127947166653\\
16.2334982557627	0.843495173159724\\
16.400326461884	0.849633711538753\\
16.5671546680053	0.855550016936354\\
16.7339828741265	0.861250512383771\\
16.9008110802478	0.86674157407575\\
17.0676392863691	0.872029517763141\\
17.2344674924904	0.877120586663066\\
17.4012956986117	0.882020940780705\\
17.568123904733	0.886736647541149\\
17.7349521108542	0.891273673634322\\
17.9017803169755	0.895637877980618\\
18.0686085230968	0.89983500572956\\
18.2354367292181	0.903870683208453\\
18.4022649353394	0.90775041374259\\
18.5690931414607	0.911479574273119\\
18.735921347582	0.915063412703097\\
18.9027495537032	0.918507045906536\\
19.0695777598245	0.921815458339424\\
19.2364059659458	0.924993501195732\\
19.4032341720671	0.928045892055197\\
19.5700623781884	0.930977214973453\\
19.7368905843097	0.933791920968487\\
19.9037187904309	0.936494328860836\\
20.0705469965522	0.939088626428039\\
20.2373752026735	0.94157887183689\\
20.4042034087948	0.943968995319883\\
20.5710316149161	0.946262801064871\\
};
\addlegendentry{Log-normal}

\end{axis}
\end{tikzpicture}%
    % This file was created by matlab2tikz.
%
%The latest updates can be retrieved from
%  http://www.mathworks.com/matlabcentral/fileexchange/22022-matlab2tikz-matlab2tikz
%where you can also make suggestions and rate matlab2tikz.
%
\begin{tikzpicture}

\begin{axis}[%
%width=9.253in,
%height=7.397in,
%at={(1.22in,1.249in)},
%scale only axis,
width=0.35\linewidth, 
scale only axis,
font=\footnotesize,
point meta min=0,
point meta max=1,
xmin=30,
xmax=85,
xlabel style={font=\color{white!15!black}, font=\footnotesize},
xlabel={$\sigma_\phi$ [$^\circ$]},
ymin=0,
ymax=1,
ylabel style={font=\color{white!15!black},font=\footnotesize, yshift=-1em},
ylabel={ECDF},
axis background/.style={fill=white},
xmajorgrids,
ymajorgrids,
legend style={at={(0.25,0.38)}, anchor=north west, legend cell align=left, align=left, draw=white!15!black, font=\footnotesize}
]
\addplot [color=red, line width=1.5pt]
  table[row sep=crcr]{%
34.3550223835542	0\\
34.3550223835542	0.0217391304347826\\
35.1220853566023	0.0217391304347826\\
35.1220853566023	0.0434782608695652\\
35.3761107230987	0.0434782608695652\\
35.3761107230987	0.0652173913043478\\
35.9070848338612	0.0652173913043478\\
35.9070848338612	0.0869565217391304\\
36.5469190226386	0.0869565217391304\\
36.5469190226386	0.108695652173913\\
38.9018003428599	0.108695652173913\\
38.9018003428599	0.130434782608696\\
39.9083717486357	0.130434782608696\\
39.9083717486357	0.152173913043478\\
40.0613422049959	0.152173913043478\\
40.0613422049959	0.173913043478261\\
40.084510953639	0.173913043478261\\
40.084510953639	0.195652173913043\\
40.941834442233	0.195652173913043\\
40.941834442233	0.217391304347826\\
41.1937889627101	0.217391304347826\\
41.1937889627101	0.239130434782609\\
43.8514700711409	0.239130434782609\\
43.8514700711409	0.260869565217391\\
44.7589041030092	0.260869565217391\\
44.7589041030092	0.282608695652174\\
44.9177270561623	0.282608695652174\\
44.9177270561623	0.304347826086956\\
45.1321102757695	0.304347826086956\\
45.1321102757695	0.326086956521739\\
46.3887345620817	0.326086956521739\\
46.3887345620817	0.347826086956522\\
46.4816822530677	0.347826086956522\\
46.4816822530677	0.369565217391304\\
47.9732681076861	0.369565217391304\\
47.9732681076861	0.391304347826087\\
48.0218753417211	0.391304347826087\\
48.0218753417211	0.413043478260869\\
48.5342173302735	0.413043478260869\\
48.5342173302735	0.434782608695652\\
49.498965889261	0.434782608695652\\
49.498965889261	0.456521739130434\\
49.5830852984758	0.456521739130434\\
49.5830852984758	0.478260869565217\\
49.7282455074516	0.478260869565217\\
49.7282455074516	0.5\\
50.2176492739829	0.5\\
50.2176492739829	0.521739130434782\\
52.1788858476301	0.521739130434782\\
52.1788858476301	0.543478260869565\\
52.9073669260024	0.543478260869565\\
52.9073669260024	0.565217391304347\\
57.037784346555	0.565217391304347\\
57.037784346555	0.58695652173913\\
57.3965090067785	0.58695652173913\\
57.3965090067785	0.608695652173913\\
58.8790906267648	0.608695652173913\\
58.8790906267648	0.630434782608695\\
59.1957066549139	0.630434782608695\\
59.1957066549139	0.652173913043478\\
61.5386279707972	0.652173913043478\\
61.5386279707972	0.673913043478261\\
61.7873369540103	0.673913043478261\\
61.7873369540103	0.695652173913043\\
62.2003812780075	0.695652173913043\\
62.2003812780075	0.717391304347826\\
62.2082293200282	0.717391304347826\\
62.2082293200282	0.739130434782608\\
63.7987183515726	0.739130434782608\\
63.7987183515726	0.760869565217391\\
66.7533285453584	0.760869565217391\\
66.7533285453584	0.782608695652174\\
66.8043097998425	0.782608695652174\\
66.8043097998425	0.804347826086956\\
68.6091473473083	0.804347826086956\\
68.6091473473083	0.826086956521739\\
69.6943999073658	0.826086956521739\\
69.6943999073658	0.847826086956522\\
70.2524793394346	0.847826086956522\\
70.2524793394346	0.869565217391304\\
71.4296477551868	0.869565217391304\\
71.4296477551868	0.891304347826087\\
72.376683437736	0.891304347826087\\
72.376683437736	0.91304347826087\\
74.7128726891423	0.91304347826087\\
74.7128726891423	0.934782608695652\\
78.3209988130337	0.934782608695652\\
78.3209988130337	0.956521739130435\\
81.7316340529054	0.956521739130435\\
81.7316340529054	0.978260869565217\\
94.3924572912165	0.978260869565217\\
94.3924572912165	1\\
};
\addlegendentry{ECDF}

\addplot [color=black, dashed, line width=2.0pt]
  table[row sep=crcr]{%
33.7546480344776	0.0414435696502385\\
34.3732155456474	0.0481455963265561\\
34.9917830568173	0.0555458456567531\\
35.6103505679871	0.0636634944285842\\
36.228918079157	0.0725124710508525\\
36.8474855903269	0.0821012162033686\\
37.4660531014967	0.092432539839659\\
38.0846206126666	0.103503573418923\\
38.7031881238364	0.115305813910775\\
39.3217556350063	0.127825254103587\\
39.9403231461761	0.141042592085307\\
40.558890657346	0.154933511467081\\
41.1774581685158	0.169469022982669\\
41.7960256796857	0.184615857505048\\
42.4145931908555	0.200336900250872\\
43.0331607020254	0.216591655960994\\
43.6517282131952	0.233336735113587\\
44.2702957243651	0.250526351705708\\
44.8888632355349	0.268112823788374\\
45.5074307467048	0.286047068719511\\
46.1259982578747	0.304279085970561\\
46.7445657690445	0.32275842125112\\
47.3631332802144	0.341434606670552\\
47.9817007913842	0.360257572608887\\
48.6002683025541	0.37917802789844\\
49.2188358137239	0.398147805803827\\
49.8374033248938	0.417120174116654\\
50.4559708360636	0.436050108441257\\
51.0745383472335	0.454894528432102\\
51.6931058584033	0.473612497347368\\
52.3116733695732	0.492165385805354\\
52.930240880743	0.510517001071029\\
53.5488083919129	0.52863368356191\\
54.1673759030828	0.546484372549292\\
54.7859434142526	0.564040643247671\\
55.4045109254225	0.581276717637839\\
56.0230784365923	0.598169451463746\\
56.6416459477622	0.614698299886544\\
57.260213458932	0.630845264277652\\
57.8787809701019	0.646594822592889\\
58.4973484812717	0.661933845697753\\
59.1159159924416	0.676851501915967\\
59.7344835036114	0.691339151954545\\
60.3530510147813	0.705390236224353\\
60.9716185259511	0.719000156429494\\
61.590186037121	0.732166153146199\\
62.2087535482908	0.744887180955571\\
62.8273210594607	0.757163782537567\\
63.4458885706306	0.768997962978589\\
64.0644560818004	0.780393065394033\\
64.6830235929703	0.791353648821892\\
65.3015911041401	0.801885369205278\\
65.92015861531	0.811994864151583\\
66.5387261264798	0.821689642034632\\
67.1572936376497	0.830977975894004\\
67.7758611488195	0.839868802483007\\
68.3944286599894	0.848371626723597\\
69.0129961711592	0.856496431742704\\
69.6315636823291	0.864253594589762\\
70.2501311934989	0.871653807669386\\
70.8686987046688	0.878708005865668\\
71.4872662158386	0.885427299284975\\
72.1058337270085	0.891822911501991\\
72.7244012381784	0.897906123158384\\
73.3429687493482	0.903688220734467\\
73.9615362605181	0.909180450290897\\
74.5801037716879	0.914393975959343\\
75.1986712828578	0.919339842947583\\
75.8172387940276	0.924028944815169\\
76.4358063051975	0.928471994770067\\
77.0543738163673	0.932679500734251\\
77.6729413275372	0.936661743926396\\
78.291508838707	0.940428760712531\\
78.9100763498769	0.943990327480005\\
79.5286438610467	0.947355948296481\\
80.1472113722166	0.950534845123199\\
80.7657788833864	0.953535950360498\\
81.3843463945563	0.956367901513058\\
82.0029139057262	0.95903903777249\\
82.621481416896	0.961557398325454\\
83.2400489280659	0.963930722206295\\
83.8586164392357	0.96616644952413\\
84.4771839504056	0.968271723905241\\
85.0957514615754	0.970253396002437\\
85.7143189727453	0.972118027933658\\
86.3328864839151	0.973871898522419\\
86.951453995085	0.975521009222725\\
87.5700215062548	0.977071090620665\\
88.1885890174247	0.978527609414154\\
88.8071565285945	0.97989577578103\\
89.4257240397644	0.981180551054056\\
90.0442915509342	0.982386655629213\\
90.6628590621041	0.983518577041067\\
91.281426573274	0.984580578145903\\
91.8999940844438	0.985576705359766\\
92.5185615956137	0.986510796904579\\
93.1371291067835	0.987386491021037\\
93.7556966179534	0.988207234112128\\
94.3742641291232	0.988976288785881\\
94.9928316402931	0.989696741770232\\
};
\addlegendentry{Log-normal}

\end{axis}
\end{tikzpicture}%
  
	\caption{Empirical CDFs of the delay  and angular spread for all the positions of Campaign A-top, Campaign B-middle, and Campaign C-bottom.}
 \label{fig:AS}
 \label{fig:Spread} 
\end{center}
\end{figure}
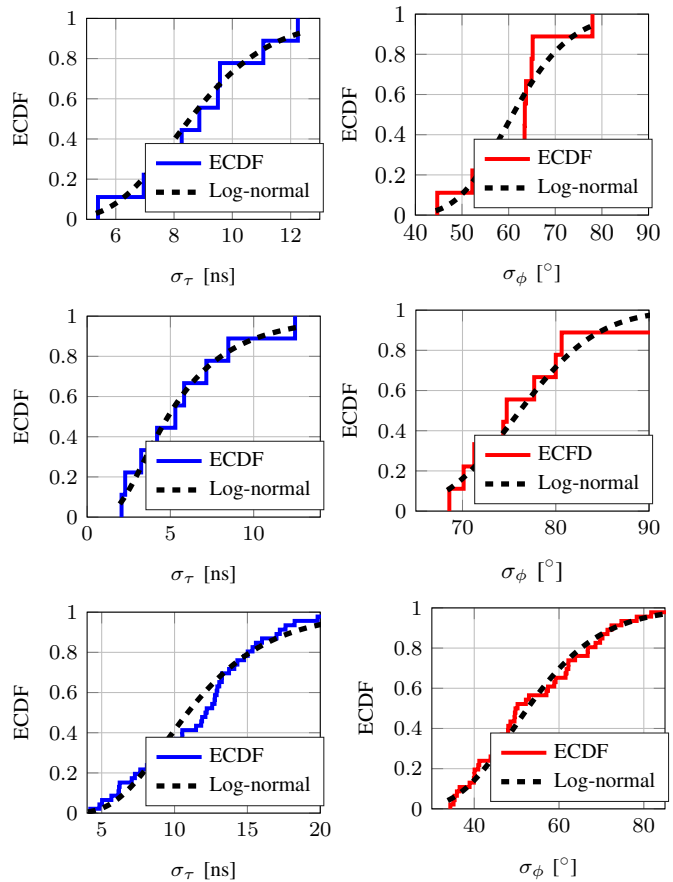

In order to characterize the backscatter channel, it is of interest to investigate the RMS delay and angular spread from the paths detected after the \ac{GEM} algorithm. The RMS delay spread is  defined as: 
\begin{equation}
    \tau_{rms}=\sqrt{\frac{\sum_{k=1}^{K} (\tau_k- \tau_m)^2 \alpha_k^2}{\sum_{k=1}^{K} \alpha_k^2}}
\end{equation}
where $\tau_m$ is the mean arrival delay defined as 
\begin{equation}
    \tau_m = \frac{\sum_{k=1}^{K} \tau_k  \alpha_k^2}{\sum_{k=1}^{K} \alpha_k^2}
\end{equation}
and $K$, $\alpha_k^2$, $\tau_k$ are the number of detected paths, the power, and the arrival delay of the $k$th path, respectively. 
The angular spread can be computed using the Fleury's definition, as in \cite{Fle:J00}
\begin{equation}
    \sigma_\phi = \sqrt{\frac{\sum_{k=1}^{K} | e^{j \phi_k} - \mu_\phi |^2 \alpha_k^2}{\sum_{k=1}^{K} \alpha_k^2}} 
\end{equation}
where  $\phi_k$ is the arrival angle of the $k$th path and $\mu_\phi$ is the mean arrival azimuth defined as 
\begin{equation}
    \mu_\phi = \frac{\sum_{k=1}^{K} e^{j \phi_k}  \alpha_k^2}{\sum_{k=1}^{K} \alpha_k^2}.
\end{equation} 
Fig.~\ref{fig:Spread} shows the empirical cumulative distribution functions (ECDFs) of  the RMS delay and the angular spread of the paths detected after the \ac{GEM} algorithm for the scenarios depicted in Fig.~\ref{fig:CampaignA}-top, \ref{fig:CampaignB}-top, and \ref{fig:CampaignC}-top. The delay spread RMSs are $8.79\,\mathrm{ns}$, $5.67\,\mathrm{ns}$, and $10.86\,\mathrm{ns}$, while the angular spread RMSs  are $61.5^\circ$, $76.4^\circ$ and $52.6^\circ$ for the three scenarios, respectively. 
The results show that delays and angular spreads can be largely dispersed according to the radar position and steering angle, and can be modeled by a log-normal distribution. 
The measured data here presented have been further used in the next section for R-SLAM purposes.
For the reader's convenience, the settings and parameters values adopted during the measurement campaigns are summarized in Table~\ref{Tab:ParametersSettings}. 
One can notice that the maximum delay spread is typically twice the one measured in the classical one-way channel in similar environment \cite{LotCaiDer:C22} which is in line with the double convolution of the back-scattering channel impulse response.

%\begin{figure}[t]
%\begin{center}
%\centering
    %{\includegraphics[width=0.48\linewidth,draft=false]{Figures/DS_side2.eps}}    {\includegraphics[width=0.48\linewidth,draft=false]{Figures/AS_side2.eps}}
    %{\includegraphics[width=0.48\linewidth,draft=false]{Figures/DS_side3.eps}}
    %{\includegraphics[width=0.48\linewidth,draft=false]{Figures/AS_side3.eps}}
    %{\includegraphics[width=0.48\linewidth,draft=false]{Figures/DS_v2.eps}}
    %{\includegraphics[width=0.48\linewidth,draft=false]{Figures/AS_v2.eps}}\label{fig:AS}
	%\caption{Delay spread and angular spread for all the positions of Campaign A (a)-(b), Campaign B (c)-(d), and Campaign C (e)-(f). \red{Ingrandire il font}}\label{fig:Spread} 
%\end{center}
%\end{figure}

\begin{table}[t] 
  \begin{center}
    \caption{Measurement campaign: parameters and settings}
    \label{tab:table1}
    \begin{tabular}{|p{4cm}|p{1.9cm}|p{1.7cm}|}
    \hline
      \textbf{Description} & \textbf{Parameter} & \textbf{Value}\\
       \hline
      \hline
            \multicolumn{3}{|c|}{Measurement setup}\\ % <-- Combining two cells with alignment c| and content 12.
       Frequency sweep range & -- &235-320 $\mathrm{GHz}$\\
       \hline
       Frequency sweep step & -- & 10 MHz\\
       \hline
        Number of \ac{CIR} samples & $M$ &8501\\
        \hline
       Angular range & -- & $[-90^{\circ}$, $90^{\circ}]$\\
         \hline
       Angular steering step & -- & $1^\circ$\\
       \hline
       Number of steering angles & $N$ &181\\
      \hline
       Transmit antenna gain & $G_{\text{TX}}$ & $20\, \text{dBi}$\\
            \hline
      Receive antenna gain & $G_{\text{RX}}$ & $20\, \text{dBi}$\\
            \hline
      Half Power Beam Width & HPBW & $18^\circ$\\
    \hline
       Detection threshold & -- & -100 dB\\
    \hline
      Delay resolution & ${T_{\text{s}}}$  &  1.56 ps\\
    \hline
      \multicolumn{3}{|c|}{Measurement data processing}\\ % <-- Combining two cells with alignment c| and content 12.
      \hline
      \ac{NM} algorithm threshold & $\eta_{\text{CF}}$ & $10^{-2}$\\
            \hline
      \ac{GEM} algorithm threshold & $\eta_{\text{CL}}$ & 0.4\\
            \hline
      Scan vector threshold  & $\eta_{\text{SV}}$ & 0.9\\
      \hline
      \multicolumn{3}{|c|}{Kalman filter parameters}\\ % <-- Combining two cells with alignment c| and content 12.
       \hline
      Power Spectral Density (PSD) of the linear acceleration noise & $w_0$ &  $10^{-4}$\\
      \hline
      PSD of the angular acceleration noise & $w_{\theta}$ & $10^{-4}$ \\
      \hline
      Time step & $T_{\text{F}}$ & 1 s \\
      \hline
       Position estimation error std & $\sigma_x=\sigma_y$ & $4.7\cdot 10^{-3}\,$m \\
       \hline
        Angle estimation error std& $\sigma_{\theta}$ & $1.7\cdot 10^{-3}\,$rad \\
    \hline
    \end{tabular}
    \label{Tab:ParametersSettings}
  \end{center}
\end{table}

\section{R-SLAM Experimental Performance}
\label{sec:slamperformance}

In this section, we investigate the performance of the \ac{R-SLAM} algorithms  using the radar measurements as input. Since the state $\textbf{x}_{k}$  of the radar at instant $k$ is estimated starting from relative pose estimations,  the final estimated trajectory and map will be relative to its initial position, which for convenience we set at the origin of the coordinate system. %, i.e., $x_0=y_0=0$. 

\begin{figure}
\centering 
\quad\quad\includegraphics[width=0.75\linewidth,draft=false]{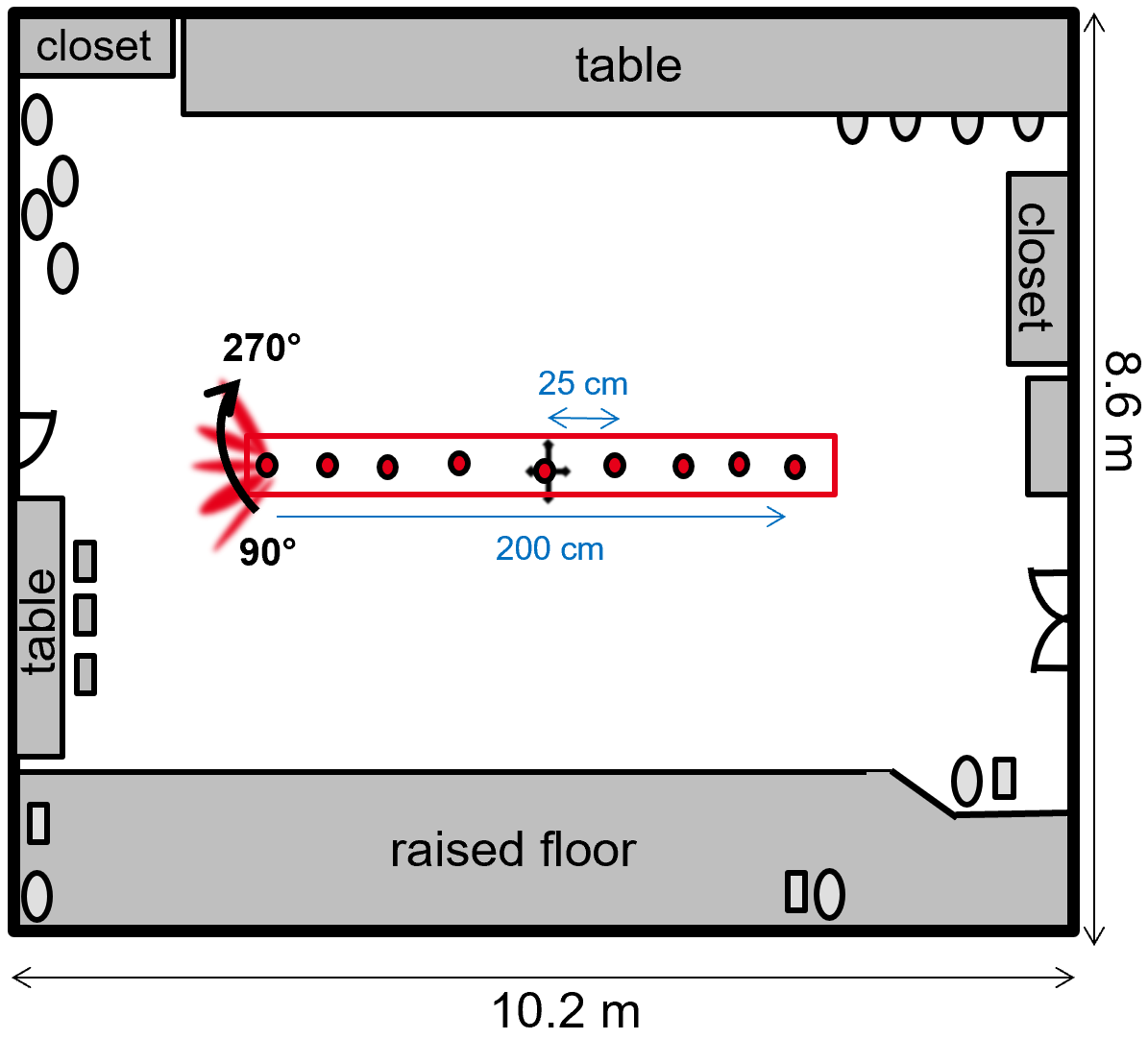}
\includegraphics[width=0.8\linewidth,draft=false]{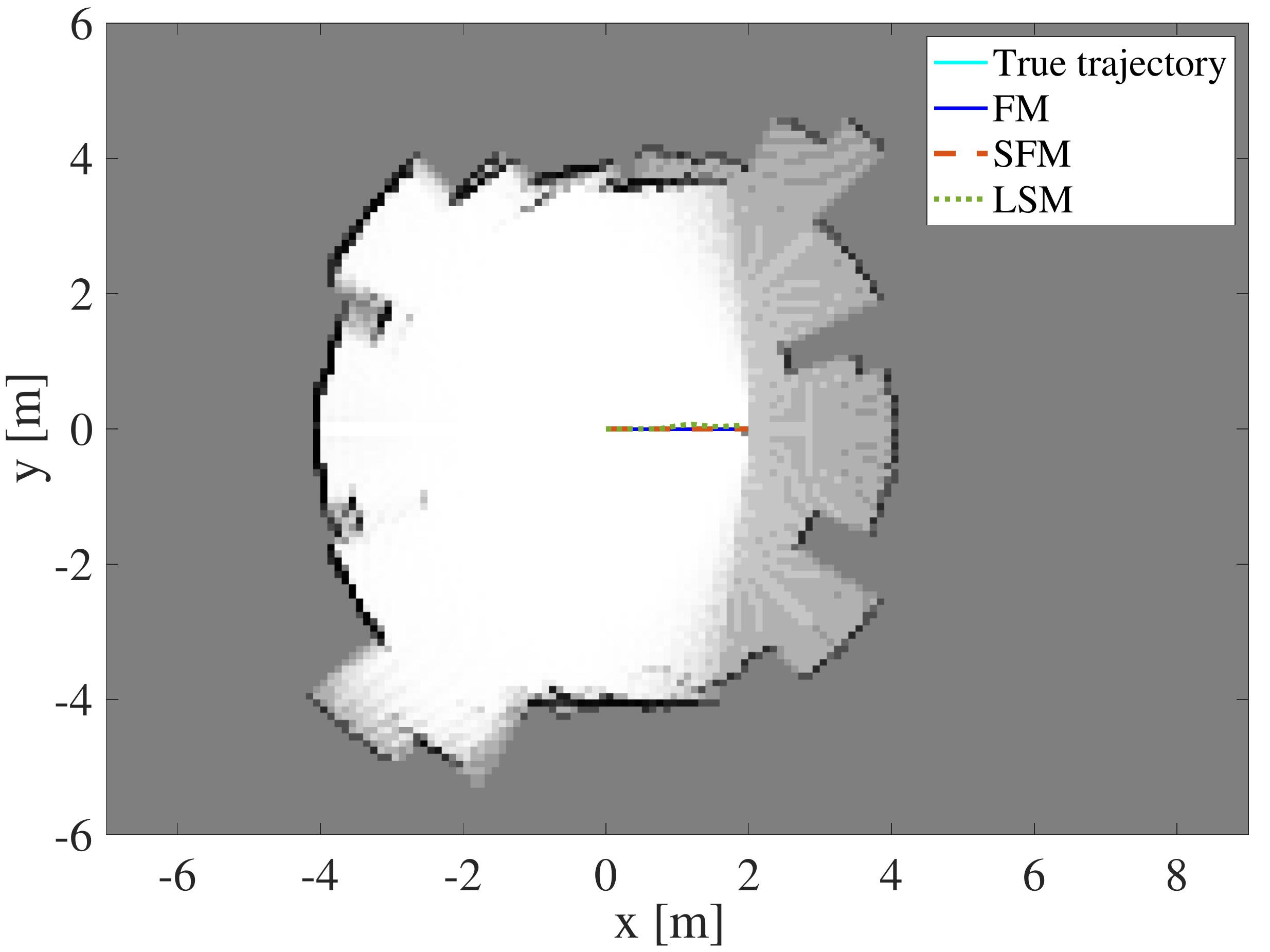}
\caption{Measurement Scenario A (top); Estimated trajectories and map (bottom). }\label{fig:CampaignA}
\end{figure}

\begin{figure}
\centering 
\quad\quad\includegraphics[width=0.75\linewidth,draft=false]{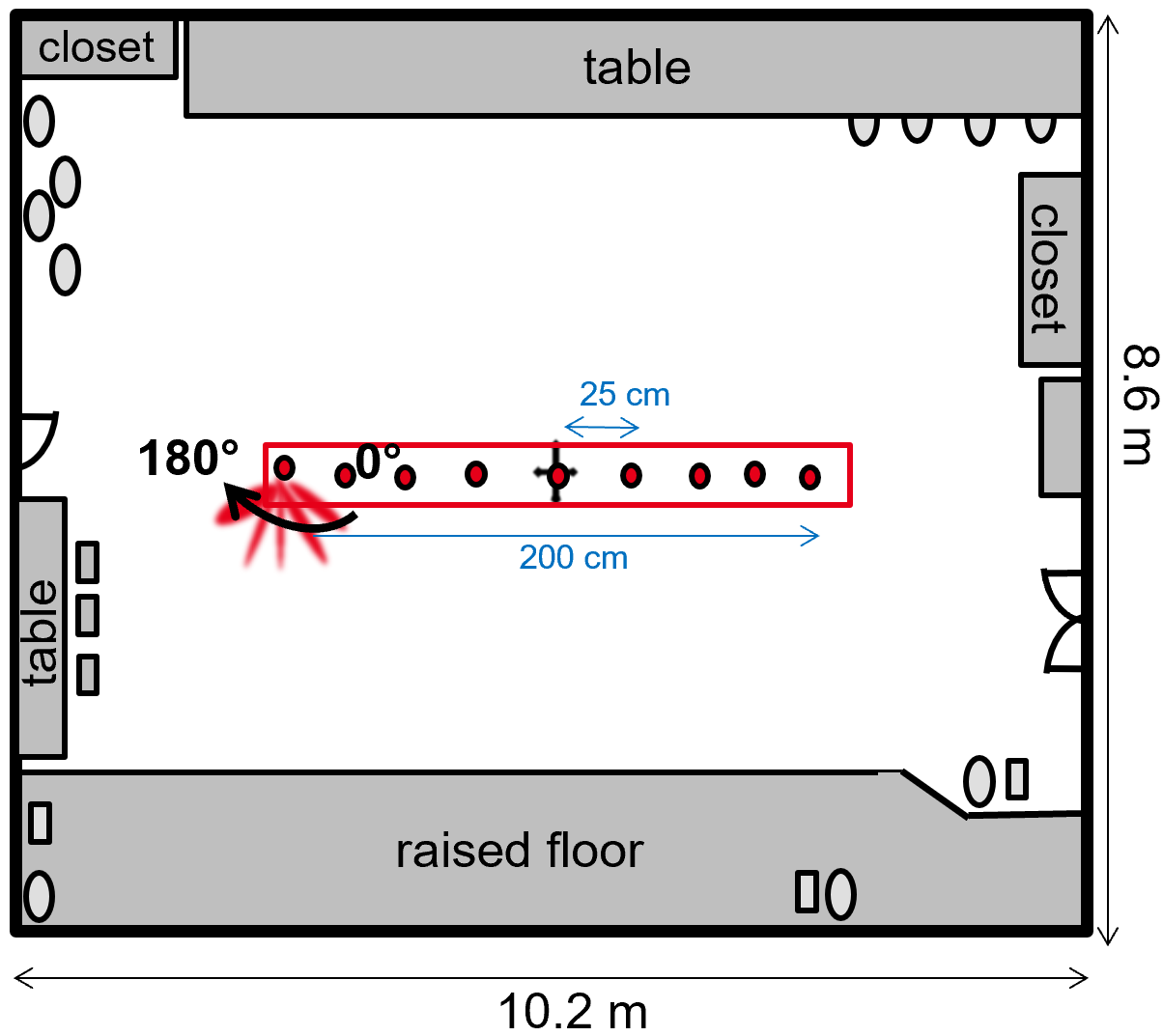}
\includegraphics[width=0.8\linewidth,draft=false]{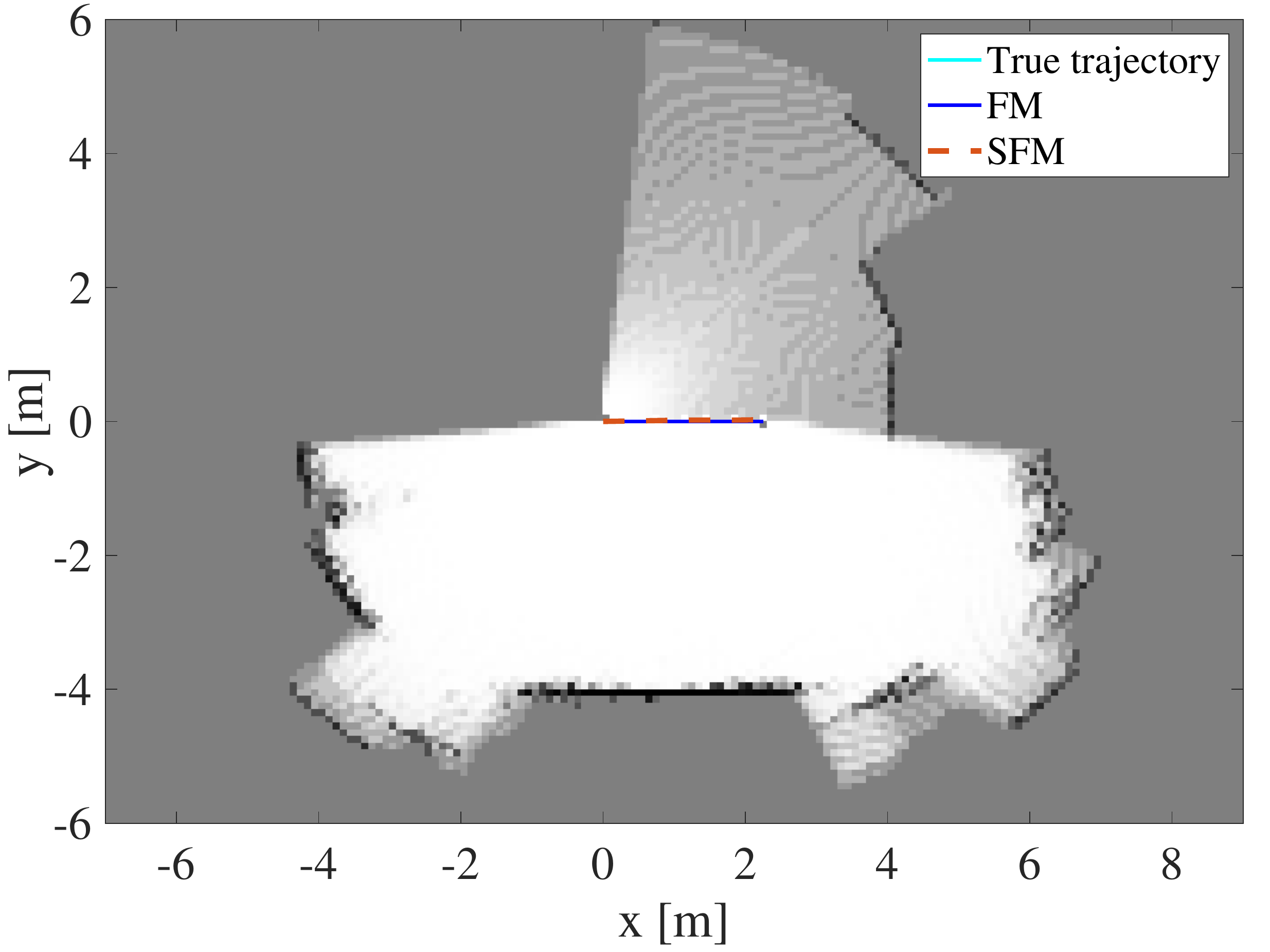}
\caption{Measurement Scenario B (top); Estimated trajectories and map (bottom).  }\label{fig:CampaignB}
\end{figure}

\begin{figure}
\centering 
\quad\quad\includegraphics[width=0.75\linewidth,draft=false]{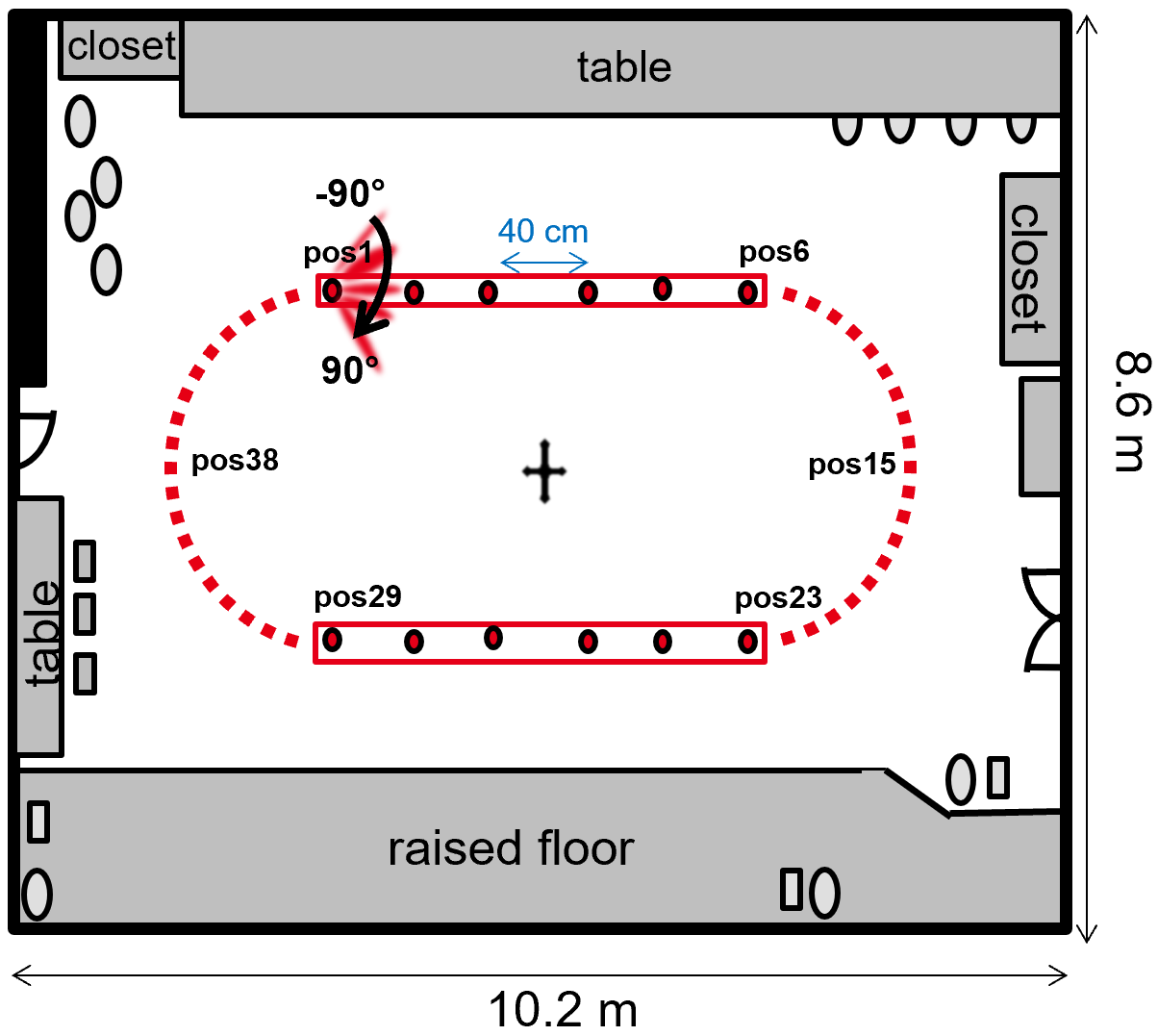}
\includegraphics[width=0.8\linewidth,draft=false]{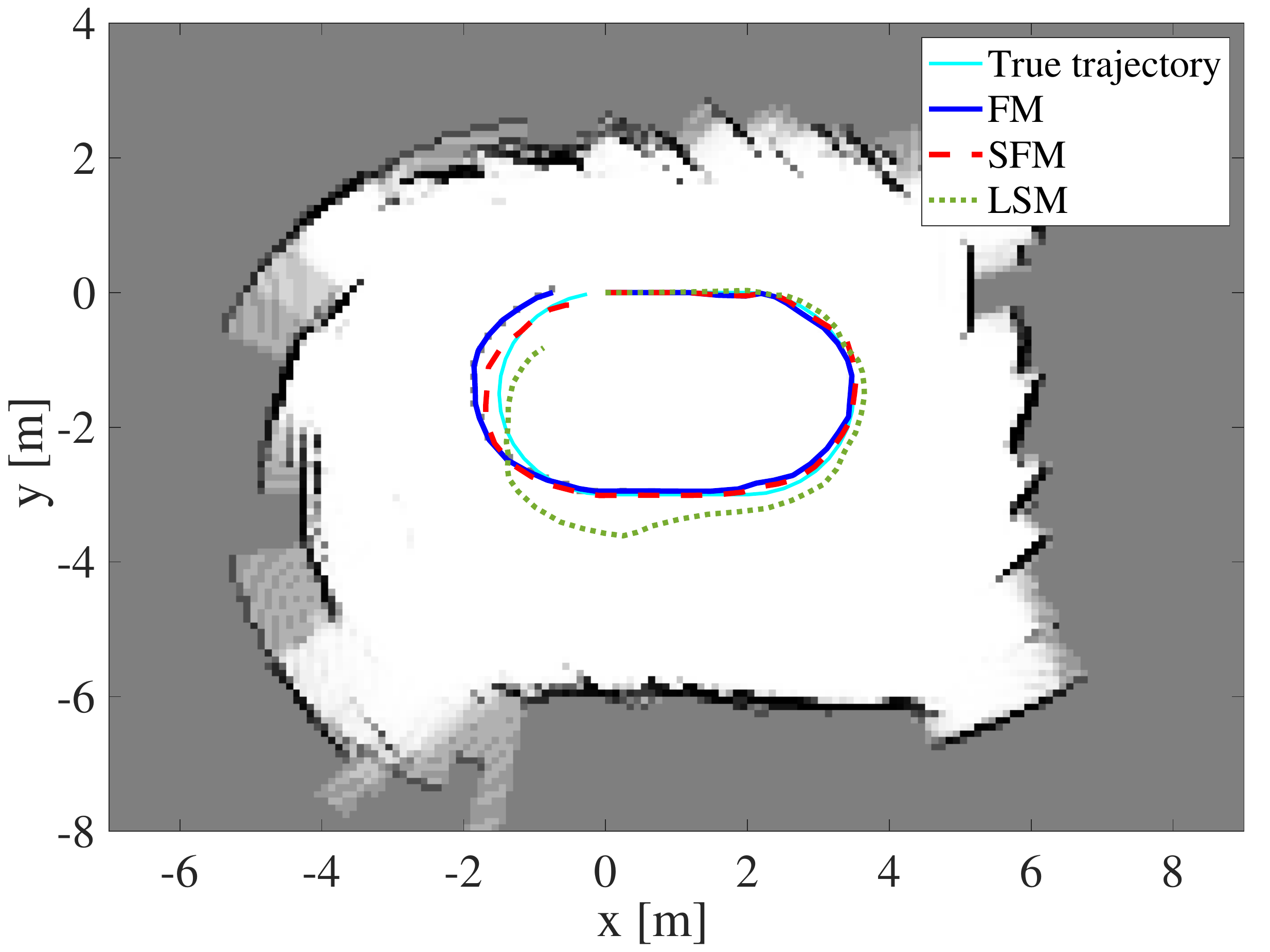}
\caption{Measurement Scenario C (top); Estimated trajectories and map (bottom).}\label{fig:CampaignC}
\end{figure}

The lower parts of Figs.~\ref{fig:CampaignA},~\ref{fig:CampaignB}, and~\ref{fig:CampaignC} show the estimated trajectories obtained from measurements taken, respectively, in the scenarios depicted in the upper parts of Figs.~\ref{fig:CampaignA},~\ref{fig:CampaignB},~\ref{fig:CampaignC} and compared with the ground-truth trajectory (cyan curve).
 Specifically, the trajectories have been estimated using the {\em Fourier-Mellin-based} (FM), {\em Simplified Fourier-Mellin} (SFM), and the \emph{Laser Scan Matching} (LSM) algorithms for the relative pose estimation. A summary of the trajectory estimation \ac{RMSE} for different approaches and scenarios is reported in Tab.~\ref{tab:rmse}.
 All algorithms provide the best trajectory estimation performance in Scenario A of Fig.~\ref{fig:CampaignA} (the corresponding curves are overlapped), with a  \ac{RMSE} of  $5.7\,\mathrm{mm}$ (FM), $5\,\mathrm{mm}$ (SFM), and $6.5\,\mathrm{cm}$ (LSM). According to the achieved results, the millimeter-level accuracy achieved with FM and SFM is of great interest.
 
 Somewhat more challenging is Scenario B in Fig.~\ref{fig:CampaignB}, where the same trajectory is followed while keeping the radar pointed perpendicular to the direction of movement.
 In this case, the relative pose relies mainly on \ac{CIR} variations in the angular domain that might be difficult to be recognized because variations in the incident angle of the electromagnetic waves on walls typically correspond to significant backscatter intensity fluctuations. As a consequence, the  \acp{RMSE} are slightly higher than those obtained in Scenario A, that  is $15\,\mathrm{cm}$ (FM), $15\,\mathrm{cm}$ (SFM), and $3.3\,\mathrm{cm}$ (LSM), denoting a higher sensitivity of FM and SFM algorithms to this effect.
 
 Finally, a more elaborated trajectory following an oval path is considered in Scenario C (Fig.~\ref{fig:CampaignC}), where simultaneous translations and rotations are present.
The corresponding trajectory estimation \acp{RMSE} are  $24\,\mathrm{cm}$ (FM), $12\,\mathrm{cm}$ (SFM), and $74\,\mathrm{cm}$ (LSM).
Surprisingly, the best performance is obtained with the SFM algorithm, whereas the LSM results in a relatively high estimation error.
This can be attributed to the fact that although FM should theoretically provide the best estimate of the relative pose, this is only true when dealing with perfect images, that is, images that are exactly a rotated and translated version of each other.
Unfortunately, radar images are far to be perfect because of multipath, noise, and the above-mentioned backscatter intensity fluctuations that create several artifacts.
Since FM and the LSM algorithms involve much more processing steps that SFM, the latter results to be more robust to artifacts.

As far as the mapping is regarded, the color of the estimated map represents the occupancy status: black, gray, and white cells indicate the value of the belief at the end of the mapping process, being equal to $1$ (occupied), $0.5$ (complete uncertainty), or $0$ (empty),  respectively. Mapping has been obtained starting from the FM trajectory estimate. In fact, since all algorithms achieved cm-level accuracies, the maps obtained with SFM and LSM show no significant differences visually. As evident, the estimated maps are consistent with the actual shape of the considered scenario.

%QUESTO PARAGRAFO ERA RIPETUTO, HO COMMENTATO IL SECONDO: As for mapping, the color of the estimated map represents the occupancy status: black, gray and white cells indicate the belief value at the end of the mapping process, which is $1$ (occupied), $0.5$ (complete uncertainty) or $0$ (empty), respectively. The mapping was obtained from the FM estimated trajectory. It is evident that the estimated maps are consistent with the actual shape of the considered scenario.

% As can be seen, for all three measurement campaigns there is a good match,  with a global positioning error in the order of 10-20 cm for the oval path and even less for the linear paths. 

These results prove experimentally the feasibility of \ac{R-SLAM} using backscattered signals in the THz band collected by a mobile radar and the possibility to achieve cm-level localization accuracy without exploiting any dedicated infrastructure.   

\begin{table}[t!]
\caption{Trajectory estimation \ac{RMSE} for different approaches and scenarios.}
\label{tab:rmse}
\centering
\begin{tabular}{c|c|c|c|} 
\multicolumn{1}{c|}{}   \cellcolor{purple!25} & Scenario A &  \cellcolor{purple!25} Scenario B & \cellcolor{purple!25} Scenario C \\
\hline
\cellcolor{green!25} FM & $5.7\,\mathrm{mm}$ & $15\,\mathrm{cm}$  &$24\,\mathrm{cm}$
\\ \hline 
\cellcolor{green!25}  SFM & $5\,\mathrm{mm}$ & $15\,\mathrm{cm}$ & $12\,\mathrm{cm}$ 
\\ \hline 
\cellcolor{green!25} LSM & $6.5\,\mathrm{cm}$ & $3.3\,\mathrm{cm}$ & $74\,\mathrm{cm}$
\\ \hline 
\end{tabular}
%\end{adjustbox}

\end{table}

%\begin{figure}
%\begin{center}
%\centering
%   \subfigure[]{\includegraphics[width=0.46\linewidth]{Figures/Rectangular_room_linear}\label{fig:Rectangular_room_linear}}
%     \subfigure[]{\includegraphics[trim= {95 15 115 50}, clip, width=0.48\linewidth]{Figures/LineSide3.eps}
%     \label{fig:CampaignA}}
%	\caption{Measurement Scenario A (a); Estimated trajectories and map (b). }
%\end{center}
%\end{figure}

%\begin{figure}
%\begin{center}
%\centering
%    \subfigure[]{\includegraphics[width=0.46\linewidth]{Figures/Rectangular_room_linear_s2}\label{fig:Rectangular_room_linear_s2}}
%     \subfigure[]{\includegraphics[trim= {95 15 115 50}, clip, width=0.48\linewidth]{Figures/LineSide2.eps}\label{fig:CampaignB}}
%	\caption{Measurement Scenario B (a); Estimated trajectories and map (b).  }
%\end{center}
%\end{figure}

%\begin{figure}
%\begin{center}
%\centering
%   \subfigure[]{\includegraphics[width=0.48\linewidth]{Figures/Rectangular_room_oval}\label{fig:Rectangular_room_oval}}
%     \subfigure[]{\includegraphics[
%     trim= {105 15 125 50}, clip,
 %    width=0.48\linewidth]{Figures/Oval}\label{fig:CampaignC}}
%	\caption{Measurement Scenario C (a); Estimated trajectories and map (b). }
%\end{center}
%\end{figure}

\section{Conclusion}
\label{sec:conclusions}

In this paper, we presented an \ac{R-SLAM} algorithm based on Fourier-Mellin transforms, together with its simplified version, which can operate with signals generated by a mobile radar. An ad-hoc measurement campaign was carried out in the \ac{THz} band with the aim of characterizing the \ac{THz} backscattering channel model and assessing the performance of the proposed \ac{R-SLAM} algorithms in a real-world scenario. The numerical results demonstrate the feasibility of infrastructure-less localization and mapping with a \emph{personal radar}, in the perspective of \ac{6G} systems where integrated communication, localization, and mapping capabilities will be needed. 
Future work will address the problem of absolute pose estimation and improving robustness to artifacts in radar images. To this purpose ultra-high gain antennas \cite{Koutsos22}, with  \ac{HPBW} less than $1^\circ$,  could be used to approximate a near-pencil beam behaviour and reduce the outliers and artifacts in radar measurements, in order to simplify the \ac{GEM} pre-processing.

\section*{Acknowledgment}
This work has been partially funded by the Horizon EU project TIMES (Grant Agreement Number: 101096307) and the H2020 Project LOCUS (Grant Agreement Number: 871249). %\textcolor{red}{This work was also sponsored by Theory Lab, Central Research Institute, 2012 Labs, Huawei Technologies Co.,Ltd.?}

\bibliographystyle{IEEEtran}
\bibliography{Biblio}

\end{document}